\documentclass[twocolumn,superscriptaddress,prb,preprintnumbers,letterpaper,floatfix,showpacs]{revtex4-1}
\usepackage{gensymb} 
\usepackage{amsmath, amssymb}
\usepackage{graphicx}
\usepackage{times}
\usepackage{dcolumn}
\usepackage{hyperref}
\usepackage{color}
\hypersetup{colorlinks = true,linkcolor = blue,anchorcolor =red,citecolor = blue,filecolor = red,urlcolor = red,
	pdfauthor=author}

\begin{document}

\title{Three-dimensional Sandglass Magnet with Non-Kramers ions}
\author{Yan-Xing Yang}
\author{Yao Wang}
\author{Zhao-Feng Ding}
\affiliation{State Key Laboratory of Surface Physics, Department of Physics, Fudan University, Shanghai 200438, China}
\author{A.~D. Hillier}
\affiliation{ISIS Facility, STFC Rutherford Appleton Laboratory, Chilton, Didcot, Oxfordshire, OX110QX, United Kingdom}

\author{Lei Shu}
\email{leishu@fudan.edu.cn}
\affiliation{State Key Laboratory of Surface Physics, Department of Physics, Fudan University, Shanghai 200438, China}
\affiliation{Collaborative Innovation Center of Advanced Microstructures, Nanjing 210093, China}
\affiliation{Shanghai Research Center for Quantum Sciences, Shanghai 201315, China}

\date{\today}

\begin{abstract}
	
	Magnetic susceptibility, specific heat, and muon spin relaxation ($\mu$SR) measurements have been performed on a newly synthesized three-dimensional sandglass-type lattice Tm$_3$SbO$_7$, where two inequivalent sets of non-Kramers Tm$^{3+}$ ions (Tm$^{3+}_1$ and Tm$^{3+}_2)$ show crystal electrical field effect at different temperature ranges. The existence of an ordered or a glassy state down to 0.1~K in zero field is excluded. The low-energy properties of Tm$_3$SbO$_7$ are dominated by the lowest non-Kramers quasi-doublet of $\rm Tm^{3+}_1$, and the energy splitting is regarded as an intrinsic transverse field. Therefore, the low-temperature paramagnetic phenomenon in Tm$_3$SbO$_7$ is explained by a transverse field Ising model, which is supported by the quantitative simulation of specific heat data. In addition, the perturbation from Tm$^{3+}_2$ may play an important role in accounting for the low temperature spin dynamics behavior observed by $\mu$SR.
	
\end{abstract}

\maketitle

\section{INTRODUCTION} \label{sec:intro}
In quantum materials, the concept of emergent phenomena due to strong-correlations between electrons or magnetic moments, quantum entanglement, topology, or frustration has attracted a lot of attention~\cite{Keimer2017,Tokura2017}. Such a concept has been widely applied in many interesting systems including high-temperature superconductor, topological insulator, and quantum spin liquid (QSL). QSL is a novel quantum system where the magnetic order is suppressed by quantum fluctuations even at zero temperature. Considerable efforts in searching for QSL have been seen since the concept was proposed by Anderson in 1973~\cite{Anderson1973}. Currently, QSL has become a hot topic in condensed matter physics due to its potential applications in quantum communication and computing\cite{Kitaev2006}.

However, experimental identification of a QSL remains a great challenge since one can not reach absolute zero temperature to identify a specific material's ground state. A compromise is to measure the enough low-temperature properties by using a variety of methods, including magnetic susceptibility, specific heat, muon spin relaxation ($\mu$SR), and nuclear magnetic resonance measurements to exclude the magnetic ordering or freezing~\cite{Broholm2020,Wen2019}. However, the observation that a material does not order magnetically at low temperatures may be caused by structure or chemical component disorder~\cite{Kimchi2018NC,Kimchi2018prx}, or is only simply due to a cooperative paramagnetic state~\cite{Keren2004,Ueland2006}. In addition to the absence of magnetic order, a recognized QSL material needs to satisfy several conditions including fractional excitations and long-range correlated dynamical spins. To demonstrate the existence of fractional excitations, the residual linear term in low-temperature specific heat~\cite{Ding2018,Ni2019,Satoshi2008c,Satoshi2011c} and thermal conductivity~\cite{Minoru2010kappa,Minoru2008kappa,Ni2019}, as well as the continuum magnetic excitation spectra in inelastic neutron scattering measurements~\cite{Coldea2001,Banerjee2017} are all expected. Besides, the low-temperature plateau of the muon spin relaxation rates is an evidence of persistent spin dynamics in an entangled spin system~\cite{Ding2018,Dunsiger2011,Chang2013}. 

So far numerous two-dimensional QSL candidates have been reported, among which either geometry frustration~\cite{Balents2010,ZhouYi2017} or Kitaev interactions~\cite{Kitaev2006,Kitagawa2018,Pei2020} introduces quantum fluctuations. However, promising three-dimensional (3D) QSL candidates are still rare since higher dimensionality suppresses the quantum fluctuations. Hyperkagome Na$_4$Ir$_3$O$_8$\cite{Okamoto2007,ChenGang2008,Zhou2008} and pyrochlore Pr$_2$Ir$_2$O$_7$~\cite{Nakatsuji2006,ChenGang2016,Yao2018,Ni2021} are the only two representative 3D QSL candidates and have been the subject of extensive studies. Even if the absence of magnetic order and spin dynamics are found, it is not sufficient to claim a QSL material, since those observations can be explained by other mechanism.

We report magnetic susceptibility, specific heat, and $\mu$SR studies of a newly synthesized fluorite oxide Tm$_3$SbO$_7$,  in which two sets of non-Kramers Tm$^{3+}$ ions (Tm$^{3+}_1$ and Tm$^{3+}_2)$ form a three-dimensional sandglass-type lattice. The absence of magnetic order or glassy state is confirmed down to 0.1~K in zero magnetic field.  The calculated magnetic entropy shows a two-step release, indicates the inequivalent Tm$^{3+}_1$ and Tm$^{3+}_2$ play a part in different energy scales. The crystal electric field (CEF) calculation suggests that the low-energy properties of Tm$_3$SbO$_7$ are dominated by the lowest non-Kramers quasi-doublet of $\rm Tm^{3+}_1$, and the energy splitting, which can be regarded as an intrinsic transverse field, is about $h\sim0.64$ meV.  Therefore, the low-temperature paramagnetic phenomenon in Tm$_3$SbO$_7$ can be described by the transverse field Ising model (TFIM)~\cite{Coldea2010,ShenYao2019,Cui2019,Liu2020,LiHan2020,ChenGang2019,Bitko1996},  since the exchange interactions between effective $S=1/2$ spins is very small according to the small Curie-Weiss temperature at low temperature. TFIM is further supported by the quantitatively simulation of specific heat data. In addition, the perturbation from Tm$^{3+}_2$ may play an important role in accounting for the low temperature spin dynamics behavior observed by $\mu$SR.

\section{EXPERIMENTS} \label{sec:exp}
Polycrystalline Tm$_3$SbO$_7$ and its non-magnetic analog Lu$_3$SbO$_7$ were synthesized by the solid state reaction. Stoichiometric amounts of Sb$_2$O$_3$ and $Ln_2$O$_3$ ($Ln$ = Tm or Lu) were mixed, thoroughly grounded, and heated at 1500 $^\circ$C for 7 days. Then polycrystalline samples were obtained after two additional re-grindings and heating. The single phase of the two samples was checked by powdered X-ray diffraction (XRD) measurements using a Bruker D8 advanced X-ray diffraction spectrometer ($\lambda$ = 1.5418 \AA). The Rietveld refinement of XRD data was conducted using FullProf software. DC magnetic susceptibility measurements were carried out in the temperature range from 2~K to 300~K by using a Magnetic Property Measurement System (MPMS, Quantum Design). The measurements of AC magnetic susceptibility from 0.1~K to 4~K and specific heat from 0.1~K to 300~K were carried out in a Physical Property Measurement System (PPMS, Quantum Design) equipped with dilution refrigerator. $\mu$SR measurements with temperatures from 0.07~K to 43~K and longitudinal external magnetic fields up to 0.3 T were performed on the MuSR spectrometer at ISIS Neutron and Muon Facility, STFC, Rutherford Appleton Laboratory, UK.

\section{RESULTS} \label{sec:results}

\subsection{\boldmath Crystal structure and CEF calculation} \label{sec:struc}

\begin{figure}[t]
	\begin{center}
		\includegraphics[width=\columnwidth]{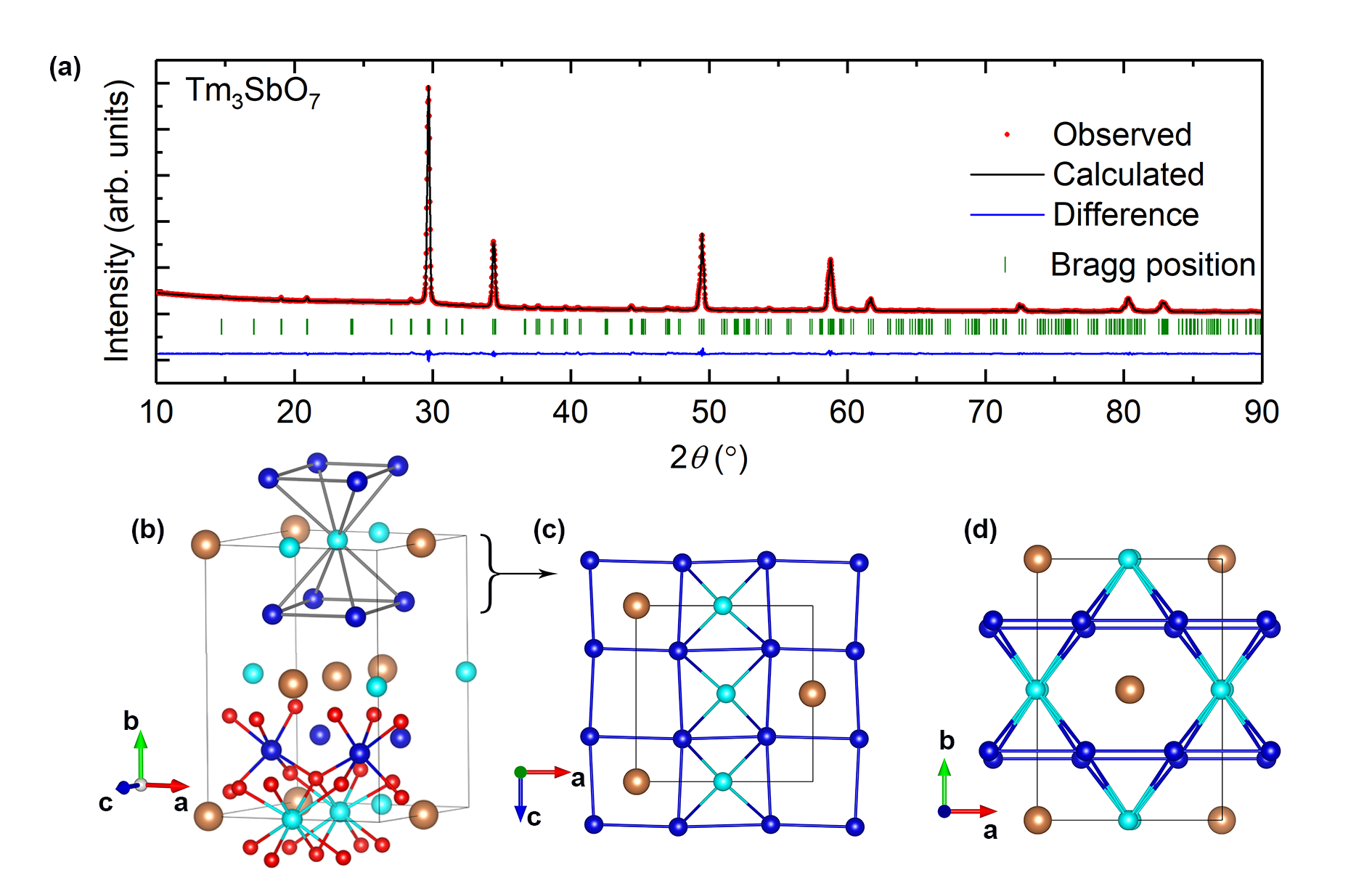}
		\caption{\textbf{(a)} Rietveld refinement of powder XRD pattern of Tm$_3$SbO$_7$ at room temperature using the orthorhombic structure with space group $C222_1$. The red dots, black line, and blue line are the experimental data, the calculated patterns, and the differences, respectively. The green bars indicates the Bragg reflections.  \textbf{(b)} The unit cell of Tm$_3$SbO$_7$ as well as the Schematic diagram of the sandglass-type configuration of Tm$^{3+}$ ions. Dark blue: Tm$_1$; light blue: Tm$_2$; brown: Sb; red: Oxygen. \textbf{(c)} Top view of two selected layers of the unit cell. \textbf{(d)} Front view of the unit cell.}
		\label{fig1}
	\end{center}
\end{figure}

The fluorite-related structure has been reported in the rare earth rhenium oxides $Ln_3$ReO$_7$ ($Ln$ = Y, Er-Lu)\cite{Masaki2018} and the authors found that $Ln_3$ReO$_7$ ($Ln$ = Y, Er, Tm) have the orthorhombic structure with space group $C222_1$ while $Ln_3$ReO$_7$ ($Ln$ = Yb, Lu) have the cubic structure with space group $Fm\bar{3}m$. The XRD pattern of Tm$_3$SbO$_7$ is shown in Fig.~\ref{fig1}(a). In the process of resolving the structure of this new material, we have firstly tried the space group $Fm\bar{3}m$ with Tm and Sb atoms randomly occupying the 4a Wyckoff position, and oxygen atoms occupying the 4b position. Only a few obvious strong reflections are eligible and these remaining weak reflections point out that the symmetry is lower than cubic structure. As a result, the orthorhombic structure with space group $C222_1$ was used, and it matches the XRD data exactly, indicating the site-mixing is not likely. The Rietveld refinement results are shown in TABLE I. 

\begin{table} [h!]
	\begin{center}
		\caption{Rietveld refinement results for Lu$_3$SbO$_{7}$.
			$R_{wp}=2.39\%, R_{p}=3.73\%, \chi^2=11.9$; 
			$a=7.389$ \AA, $b=10.398$ \AA, $c= 7.361$ \AA; 
			$\alpha=\beta=\gamma=90^{\circ}$; Space Group: $C222_1$.}
		\begin{tabular}{rcccccc}
			\hline
			\hline
			& Wyckoff \\
			Atom & positions & $x$ & $y$ & $z$ & B/\AA$^2$ & Occ.\\
			\hline
			Tm1 & 8c & 0.2424(8) & 0.2352(9) & 0.7431(6)& 0.1587(6) & 1 \\
			Tm2 & 4a & 0.0094(1) & 0.5 & 0.5& 0.1598(7) & 1 \\					
			Sb & 4a & 0.0129(8) & 0 & 0 & 0.0551(5) & 1 \\
			O1 & 8c & 0.2165(0) & 0.1351(4) & 0.4982(9) & 0 & 1 \\
			O2 & 8c & 0.2157(8) &0.1166(8) & 0.0697(4) & 4.7098(3) & 1 \\
			O3 & 4b & 0 &0.0840(5) &0.75 & 1.8571(2) & 1 \\
			O4 & 4b & 0 &0.3557(9) &0.75 & 0.0822(5) & 1 \\
			O5 & 4b & 0 &0.3640(6) &0.25 & 0.5970 & 1 \\				
			\hline
			\hline
		\end{tabular}
	\end{center}
\end{table}

The unit cell of Tm$_3$SbO$_7$ is shown in Fig.~\ref{fig1}(b). There are two different Tm sites in Tm$_3$SbO$_7$ with the ratio of Tm$_1$:Tm$_2$ = 2:1. As shown in Figs.~\ref{fig1}(b-d), Tm$_1$ atoms (dark blue dots) construct a twisted tetragonal lattice, while Tm$_2$ (light blue dots) and Sb (brown dots) atoms reside in the center of Tm$_1$ cuboids. Both Tm$_2$ and Sb atoms form the one-dimensional chains parallel to the [001] direction. As shown in Fig.~\ref{fig1}(b), a Tm$_1$ cuboid with a central Tm$_2$ atom form a sandglass-type unit. Ignoring the nonmagnetic Sb atoms, all the Tm atoms construct the edge-shared sandglass-type structure. The bond lengths of these neighboring Tm atoms vary in a small range from 3.53 \AA~to 3.87 \AA, indicating that Tm$_3$SbO$_7$ is a 3D magnet.

\begin{figure}[h]
	\begin{center}
		\includegraphics[width=4cm]{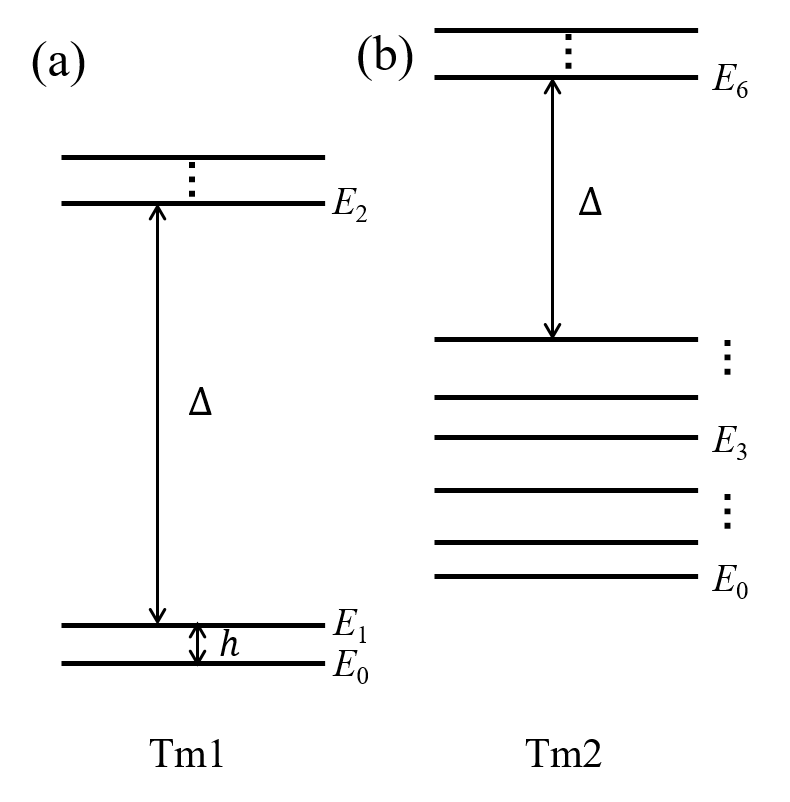}
		\caption{Schematic of the CEF splitting of \textbf{(a)} Tm$_1$ and \textbf{(b)} Tm$_2$ in Tm$_3$SbO$_7$.}
		\label{fig2}
	\end{center}
\end{figure}

The coordinate oxygen atoms of both Tm$_1$ and Tm$_2$ are presented in Fig.~\ref{fig1}(b). Tm$_1$ resides in an octahedral oxygen cavity, while Tm$_2$ resides in an 8-coordinated oxygen polyhedron. Based on the structure obtained from XRD, we did CEF calculation using the software PyCrystalField\cite{PyCrystalField}.  The 13-fold degenerate 4f orbit of each Tm$^{3+}$ will spilt due to the Coulomb potential from their surrounding ions. The intuitive schematic of CEF splitting of Tm$_1$ and Tm$_2$ is shown in Fig.~\ref{fig2}. For Tm$_1$, the ground state and the first excited state form a quasi-doublet and the energy splitting gap $h$ between the two states is about 0.25 meV, which is much smaller than $\Delta$ ($\approx$~29 meV), the energy gap between $E_1$ and $E_2$. So the low-temperature properties for Tm$_1$ is qualitatively governed by this quasi-doublet.  For Tm$_2$, The lowest 6 energy levels from $E_0$ to $E_5$ are 0 meV, 0.91 meV, 3.79 meV, 6.53 meV, 7.68 meV and 10.89 meV. They are relatively evenly distributed and gapped by a large gap ($\Delta \approx$~17 meV) from $E_6$ and higher levels. Besides, according to the specific heat results discussed in Sec.~\ref{sec: SP}, the ground state level $E_0$ of Tm$_2$ is higher than that of Tm$_1$.

\subsection{\boldmath Magnetic susceptibility} \label{sec:sus}
\begin{figure}[h]
	\begin{center}
		\includegraphics[width=7cm]{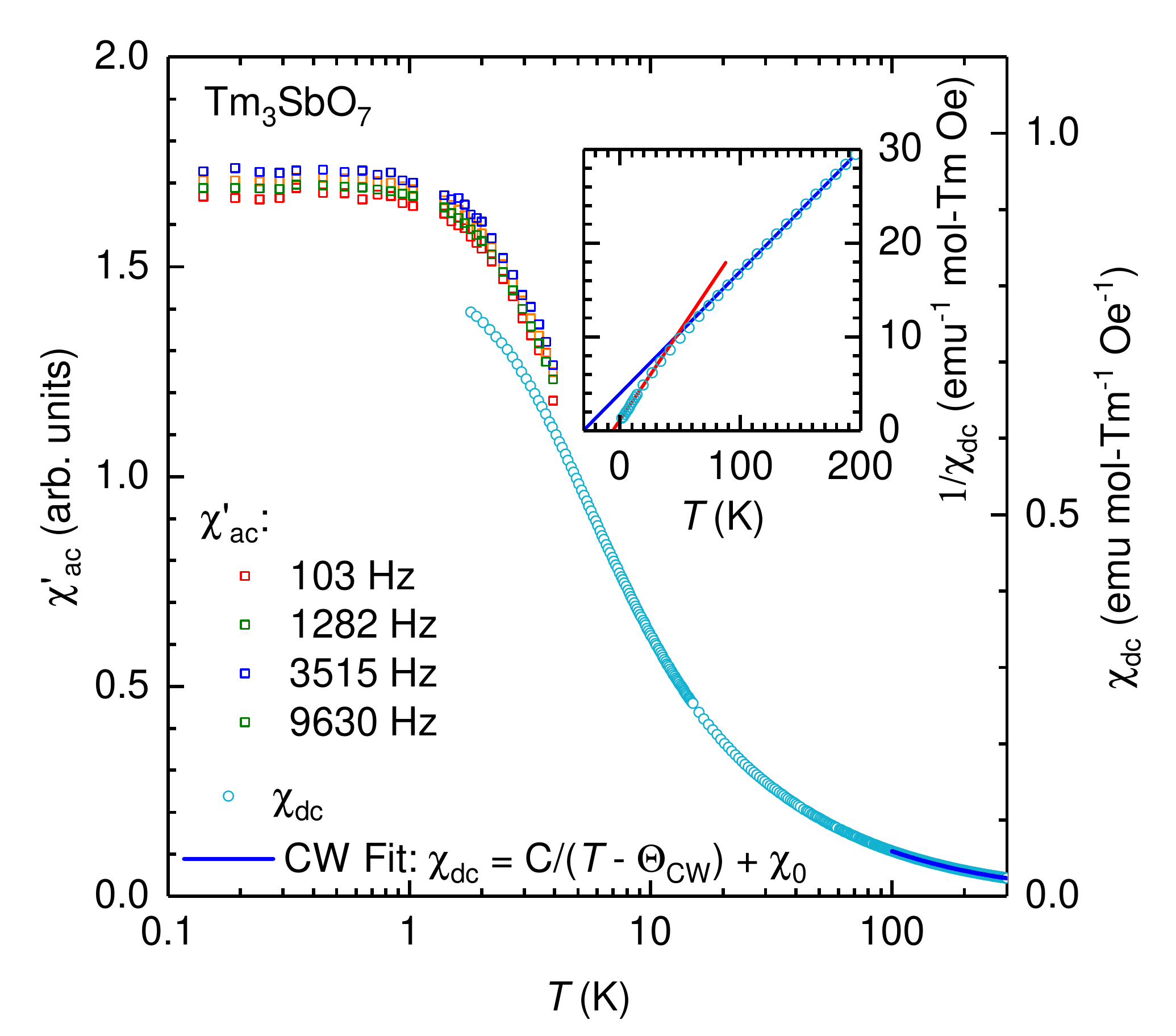}
		\caption{Temperature dependence of dc magnetic susceptibility $\chi_{dc}$ (circles) and the real part of ac susceptibility $\chi'_{ac}$ (squares). $\chi_{ac}$ was measured in zero static field with a driven field of 1 Oe from 0.1 K to 4 K. $\chi_{dc}$ was measured under $\mu_0H=0.5$~T from 2~K to 300~K. $\chi_{dc}$ between 100-300 K was fitted using Curie-Weiss law as shown in the picture. The temperature-independent $\chi_{0}$ is induced by Van-Vleck susceptibility.}  
		\label{fig3}
	\end{center}
\end{figure}

DC-magnetic susceptibility $\chi_{dc}$ of Tm$_3$SbO$_7$ measured under a magnetic field of 0.5 T from 2 K to 300 K are shown in Fig.~\ref{fig3}. No peak reflecting phase transition or separation between zero-field cooling and field cooling (not shown) is found down to 2~K.  At high temperatures, $\chi_{dc}$ increases as the temperature is reduced. The inset of Fig.~\ref{fig3} shows the inverse of $\chi_{dc}$ as function of temperature.  A fit of Curie-Weiss law is shown for temperature between 100~K and 300~K (blue line) . The Curie-Weiss temperature $\Theta_{\rm{CW}}$ is -23.3~K, and the effective magnetic moment $\mu_{\rm{eff}}$ is 7.53 $\mu_{\rm{B}}$, which is close to the theoretical value $\mu_{\rm{calc}}$ = 7.57 $\mu_{\rm{B}}$ for Tm$^{3+}$ ions with the spin-orbital coupling ground state $^3\rm{H}_6$. When the temperature is cooled down below 100 K, $\chi_{dc}$ slowly deviates from the high temperature Curie-Weiss law and forms to another Curie-Weiss behavior at low temperatures (red line), which gives $\mu_{\rm eff}=6.25~\mu_{\rm B}$ and $\Theta_{\rm{CW}}= -4.28$~K.

AC susceptibility was measured from 4 K to 0.1 K. $\chi'_{ac}$ data with different driving frequencies show a similar behavior, i.e., gradually increases with lowering the temperature and finally saturates below 1~K without showing any anomalies. Therefore, magnetic ordered state as well as the spin glass state in Tm$_3$SbO$_7$ can be ruled out.

\subsection{\boldmath Specific heat} \label{sec: SP}

\begin{figure}[t]
	\begin{center}
		\includegraphics[width=\columnwidth]{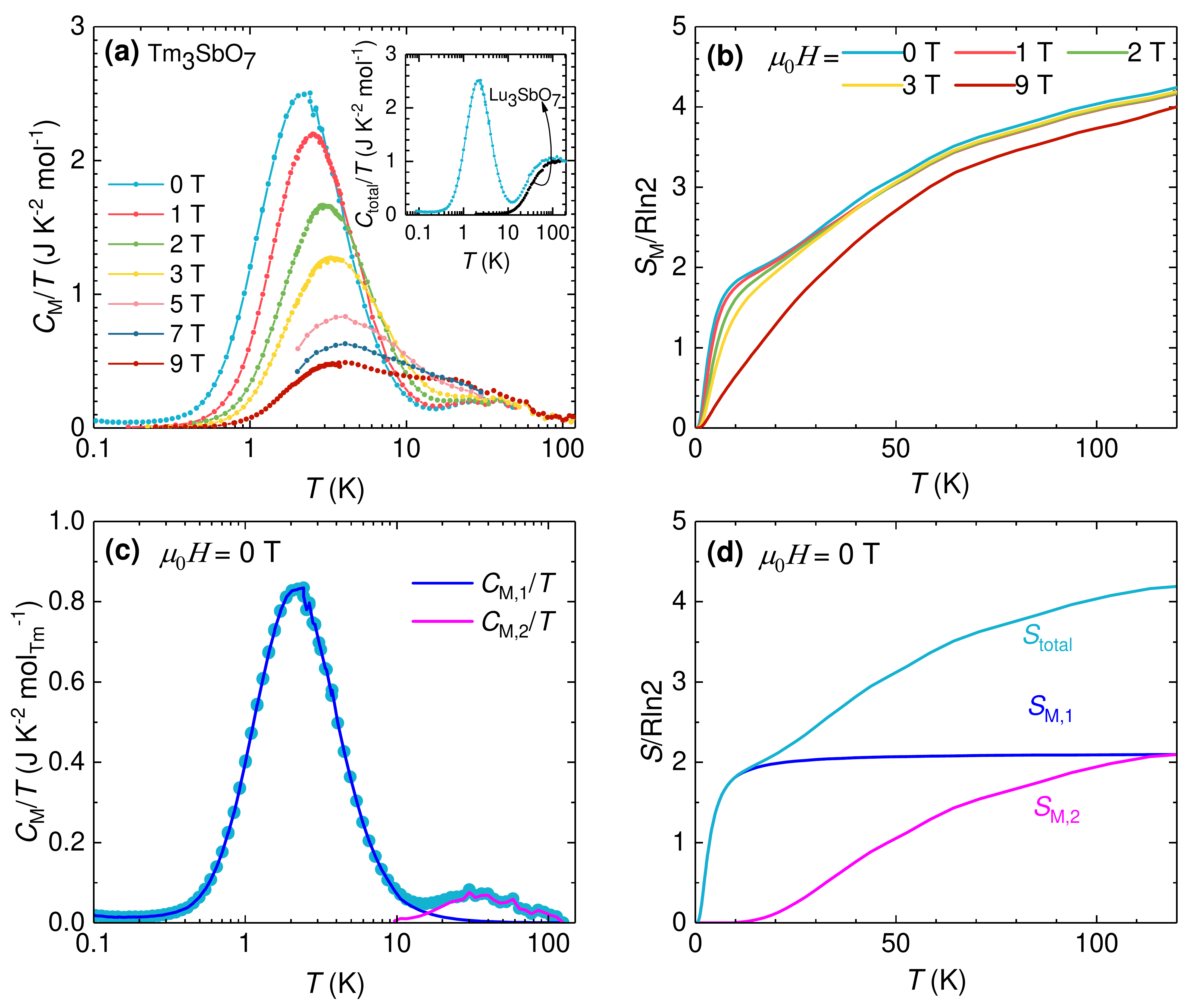}
		\caption{\textbf{(a)} Temperature dependence of magnetic specific heat coefficient $C_{\rm{M}}/T$ of Tm$_3$SbO$_7$ under several fields. Inset: Temperature dependence of measured total specific heat coefficient $C_{\rm{total}}/T$ of Tm$_3$SbO$_7$ and non-magnetic analog Lu$_3$SbO$_7$ (black points) under zero field. \textbf{(b)} Temperature dependence of magnetic entropy $S_{\rm{M}}$ obtained by integrating $C_{\rm{M}}/T$ from about $T=$ 0.1 K to $T$. \textbf{(c)} Temperature dependence of $C_{\rm{M}}/T$ of Tm$_3$SbO$_7$ under zero field. Curves are guided by eye. \textbf{(d)} Two step entropy increasing under zero magnetic field.}
		\label{fig4}
	\end{center}
\end{figure}

To further investigate the thermodynamics of Tm$_3$SbO$_7$, we measured the specific heat down to about 0.1~K by applying various magnetic fields, as shown in Fig.~\ref{fig4}(a). We subtract the phonon contribution, which is obtained from nonmagnetic oxide Lu$_3$SbO$_7$ and depicted (black points) in the inset of Fig.~\ref{fig4}(a), from the total specific heat $C_{\rm{total}}/T$ of Tm$_3$SbO$_7$. Due to the uncertainty of subtraction at high temperatures, the magnetic specific heat are only exhibited below 120~K. The whole curve of $C_{\rm{M}}/T$ shows two overlapped broad bumps. However, there are no sharp peaks throughout the full temperature range . The characteristic of no phase transition is consistent with the magnetic susceptibility above and the following $\mu$SR results. 

As displayed in Fig.~\ref{fig4}(a), by increasing the external magnetic fields, the lower-temperature bump is lowered and broadened, and the peak position moves to higher temperatures until $\mu_0H$ = 5~T. The position of the higher-temperature bump does not move but gets broader with increasing applied magnetic fields, and there is no obvious bump at $\mu_0H$ = 9~T. 

We calculated magnetic entropy $S_{\rm{M}}$ by integrating the $C_{\rm{M}}/T$ curve, as revealed in Fig.~\ref{fig4}(b). With the temperature increasing, $S_{\rm{M}}$ rises up steeply in the beginning, and then gently climbs from 10~K to 120~K.  Since the entropy under various magnetic fields have a roughly 4$R$ln2 in total for per mole sample except for a small deficiency at the highest field we measured, we believe that the energy levels below 120~K keep unchanged with magnetic field up to 9~T, which supports the CEF calculation that there is indeed a large energy gap between the lower energy levels and the much higher energy levels. As a result, in the following we only discuss the lower energy levels.

To clarify the relationship between specific heat and the CEF energy levels, we present zero-field $C_{\rm M}/T$ and $S_{\rm{M}}$ in Figs.~\ref{fig4}(c, d). As shown in Fig.~\ref{fig4}(c), $C_{\rm M}/T$ curve is roughly divided into two parts, labeled as $C_{\rm M,1}/T$ and $C_{\rm M,2}/T$, respectively. The corresponding $S_{\rm{M,1}}$ and  $S_{\rm{M,2}}$ are drawn in Fig.~\ref{fig4}(d). The CEF calculation in Sec.~\ref{sec:struc} has revealed that, for per formula unit of Tm$_3$SbO$_7$, there are 4 energy levels with small splitting for Tm$_1$  and 6 energy levels for Tm$_2$. The 4 levels of Tm$_1$ are 2-fold degenerate (two quasi-doublets) due to two equivalent Tm$_1$ atoms, while the 6 levels of Tm$_2$ are non-degenerate. Comparing the small gap $h$ between the quasi-doublet of Tm$_1$ with the gaps among Tm$_2$'s 6 levels (for instance, the 3rd excited state of Tm2 has a 6.53-meV gap from $E_{\rm 0}$, which is one-order-of-magnitude larger than $h$), we infer that $C_{\rm M,1}/T$ and $C_{\rm M,2}/T$ correspond to the contribution from Tm$_1$ and Tm$_2$, respectively.

Now we can quantitatively describe the evolution of $S_{\rm{M}}$ in Fig.~\ref{fig4}(d). For Tm$_1$, two quasi-doublets per formula unit can offer 2$R$ln2 entropy increasing, which is exactly the saturation value of $S_{\rm{M,1}}$. For Tm$_2$, as shown in Fig.~\ref{fig4}(d), $S_{\rm {M,2}}$ rises up from 10~K and reaches  $R$ln4 around 120 K, without a sign of saturation. It is reasonable that only the lowest 4 energy levels of Tm$_2$ are covered below 120~K. 

\subsection{\boldmath $\mu$SR} \label{sec: uSR}
\begin{figure}[t] 
	\begin{center}
		\includegraphics[width=\columnwidth]{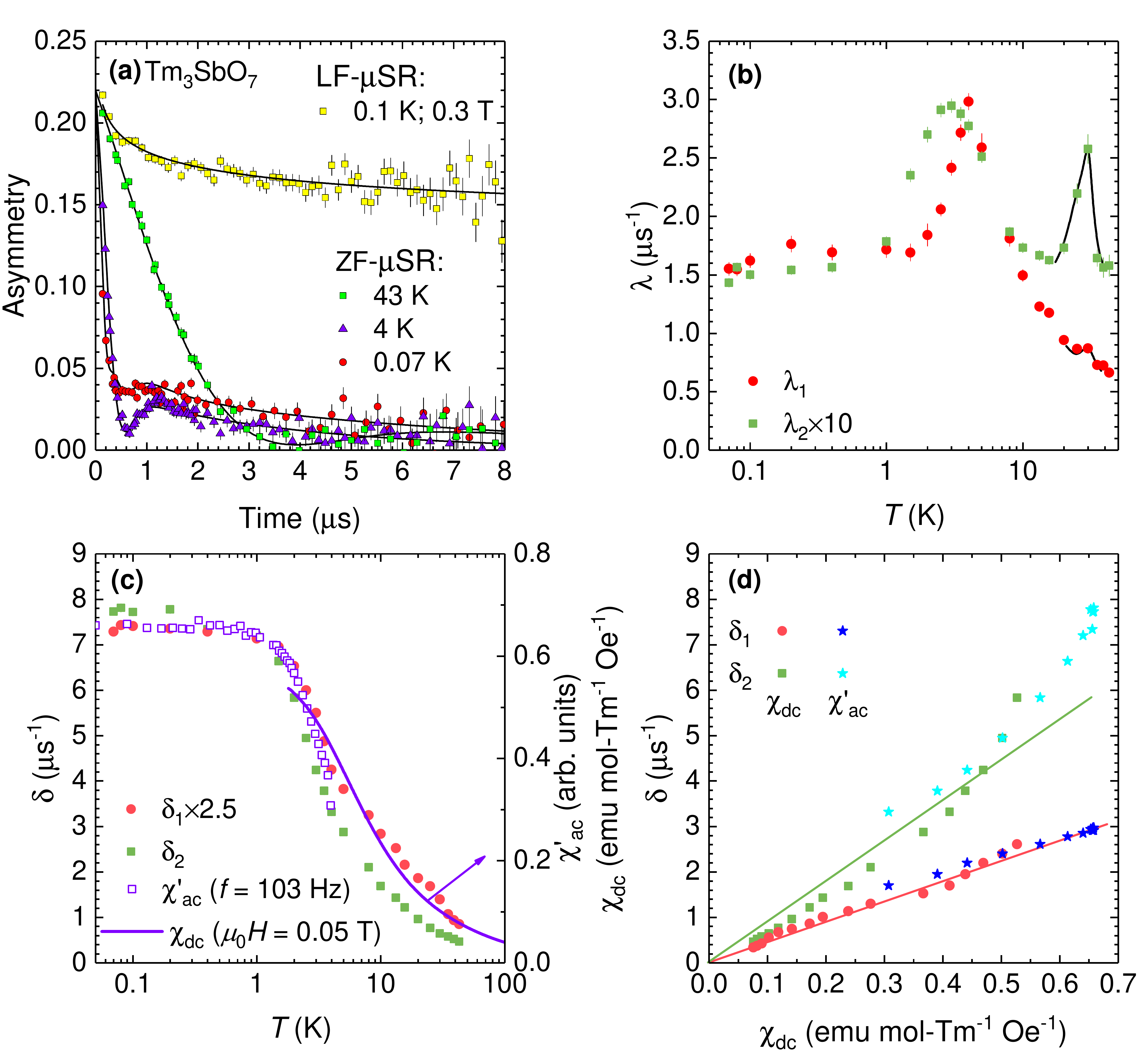}
		\caption{\textbf{(a)} Representative $\mu$SR asymmetry spectra (a constant background is subtracted) measured in ZF and LF. Solid lines are fits to the data. \textbf{(b)} Temperature dependence of ZF dynamic relaxation rate $\lambda$. The black line is to guide the eyes. \textbf{(c)} Temperature dependence of ZF static relaxation rate $\delta_1$ (red dots) and $\delta_2$ (green dots) at two different muon stopping sites. Purple line: dc-magnetic susceptibility $\chi_{\rm dc}$. Purple circle: ac-magnetic susceptibility $\chi'_{ac}$, data from Fig.~\ref{fig2} and scaled with $\chi_{\rm dc}$. \textbf{(d)} Dependence of static relaxation rate $\delta_{1,2}$ of $\chi_{dc}$ or $\chi'_{ac}$ with temperature as an implicit parameter.}
		\label{fig5}
	\end{center}
\end{figure}

$\mu$SR is a low-frequency probe of spin dynamics and is particularly sensitive to slow spin fluctuations~\cite{Hayano79}. It is therefore ideally suited to studies of long-lived spin correlations in novel spin systems. We continue to study the intrinsic magnetic properties of Tm$_3$SbO$_7$ by performing $\mu$SR experiments. Both the zero-field (ZF)- and longitudinal-field (LF)-$\mu$SR spectra are shown in Fig.~\ref{fig5}(a). The ZF-$\mu$SR asymmetries are well fitted by a sum of two damped Kubo-Toyabe functions originating from two inequivalent muon sites:
\begin{equation}
\label{AsyZF}
A(t)=A_{0} f_1 e^{-\lambda_{1} t} G_z^{\rm{K T}}(\delta_{1}, t)+
A_{0} (1-f_1) e^{-\lambda_{2} t} G_z^{\rm{K T}}(\delta_{2}, t)     
\end{equation}
where $A_0$ is the initial asymmetry and $f_1$ represents the fraction of the first muon sites in sample. During the data processing, $A_0$ and $f_1$ were found to be temperature independent and therefore are fixed at the average values of 0.22 and 0.5, respectively. The exponential rates $\lambda_{1, 2}$ are the muon spin relaxation rates usually related to the dynamic internal magnetic fields. $G_z^{\rm{K T}}(\delta, t)$ is the well-known Kubo-Toyabe function in which the relaxation rates $\delta_{1, 2}$ originate from static internal magnetic fields such as nuclear dipolar fields~\cite{Hayano79}. As shown in Fig.~\ref{fig5}(b) and (c), muons at two different stopping sites sense the same internal magnetic fields and the only difference is the strength. This indicates that two muon sites are reasonable and the phase separation can be excluded. 

Cooling from high temperatures, the dynamical relaxation rate $\lambda_{1, 2}$ gradually goes up and shows broad peaks around 3~K and finally saturates below 1~K. The low-temperature plateau of muon relaxation rates is a sign of persistent spin dynamics~\cite{Dunsiger2011,Ding2018,Chang2013}. The dynamical property of internal fields is further confirmed by the LF-$\mu$SR results (see Fig.~\ref{fig5}(a)), since the muon depolarization would be completely decoupled under such an external longitudinal field if the internal fields are static or quasi-static. Note that although $\lambda_{2}$ is ten times smaller than $\lambda_{1}$, both $\lambda_{1}$ and $\lambda_{2}$ show two peaks around the same temperatures. The two peaks of $\lambda_{1, 2}$ are also consistent with the two bumps discovered in the specific heat measurements. 

As shown in Fig.~\ref{fig5}(c), the static muon spin relaxation rates $\delta_{1, 2}$, similar to $\chi_{dc}$ and $\chi'_{ac}$, increases with decreasing temperature and saturates below 1~K. We argue that the temperature-dependent $\delta_{1, 2}$ is related to hyperfine-enhanced Tm nuclear moments~\cite{MacLaughlin2009}. The enhanced value of the static Tm nuclear contribution, $\delta$, has the formula:
\begin{equation}
\label{hyperfine}
\delta = (1+k)\delta_{0}
\end{equation}
where $\delta_{0}$ is the un-enhanced value of the static Tm nuclear contribution. $k=\rm{a_{4f}}\chi_{\rm{mol}}$ is the enhancement factor, where $\rm{a_{4f}}$ is the atomic hyperfine coupling constant and $\chi_{\rm{mol}}$ is the DC magnetic susceptibility. By replacing $k$ in Eq.~(\ref{hyperfine}) with $\rm{a_{4f}}\chi_{\rm{mol}}$, we obtain $d\delta/d{\chi_{\rm{mol}}}=\rm{a_{4f}}\delta_{0}$, indicating that $\delta$ should be proportional to $\chi_{\rm{mol}}$ with temperature as an implicit parameter. This is consistent with our experimental results as shown in Fig.~\ref{fig5}(d). $\delta_{1, 2}$ is proportional to $\chi_{dc}$ in a wide temperature range.

\section{DISCUSSION} \label{sec:discussion}

\subsection{\boldmath Dynamical muon spin relaxation rate, specific heat, and magnetic susceptibility} \label{sec:lambdagammachi}
If we compare the temperature dependence of $C_{\rm{M}}/T$ with the dynamical muon relaxation rate $\lambda_{1,2}$, we find that the specific-heat bump is consistent with the relaxation rate peak. When $T << h$, the Tm electrons tend to stay in the lowest level and the transition probability between different levels is low. When $T$ is close to $h$, the probability gets sufficient and hence leads to a maximum in $C_{\rm{M}}/T$. When $T >> h$, the probability drops again because different levels are almost equally occupied. The temperature dependence of electron transitions explains the change in specific heat. Meanwhile, since Tm$^{3+}$ ions are magnetic, the electron transitions sensed by muons are considered as magnetic fluctuations. This is why the two different methods possess some common features in the temperature dependence.

The low-temperature plateau of $\lambda$ below 1~K is a sign of persistent spin dynamics. The spin dynamics is confirmed by LF-$\mu$SR experiment.  We speculate that the perturbation from Tm$_2$ plays an important role and brings about the dynamics. When cooling the temperature across $T_s$ = 1~K, Tm$_3$SbO$_7$ may evolves into quantum paramagnetic phase. In addition, we notice that both $\chi'_{ac}$ and $\chi_{dc}$ ($\propto \delta$) also saturate under 1~K. The fine uniformity indicates what we observed are all intrinsic.

\subsection{\boldmath Transverse field Ising model} \label{sec:TFIM}

The low-temperature paramagnetic natrue of $\rm Tm_3SbO_7$ drives us to focus on the low-energy physics, especially below 10~K. At low temperatures, we can construct an effective model to describe the quasi-doublet of Tm$_1$. In general, considering the non-Kramers nature and interactions between Tm$_1$ ions, we can model the low-energy physics by using the transverse field Ising model (TFIM), which is well-studied and successfully applied in many real materials\cite{Coldea2010,ChenGang2019,Cui2019,LiHan2020,Bitko1996}. 
As metioned in \ref{sec:struc}, Tm1 forms a 3D tetragonal lattice. The Hamiltonian can be generally written as
\begin{equation}
	H=\frac{1}{2}\sum_{i j}J_{i j} S_{i}^{z} S_{j}^{z}-\sum_{i} h S_{i}^{x}
	\end{equation}\label{eq:TIFM}
where the two energy levels in the quasi-doublet of Tm$_1$ are regarded as the up and down degrees of an Ising spin $S_i$. $J_{i j}$ is the interaction energy between two spins, and $h$ is the energy splitting between the quasi-doublet of Tm$_1$ ions, acting as the intrinsic transverse field~\cite{Coldea2010,ShenYao2019,Cui2019,Liu2020,LiHan2020,ChenGang2019,Bitko1996}.

Since a 3D tetragonal lattice hardly has geometrical frustration, if $J_{i j}$ is dominant, the system should become ordered at a finite temperature, while a leading $h$ will prevent the system from ordering even at zero temperature. The energy splitting $h$ between the quasi-doublet is about 0.64 meV, whose order of magnitude is agreeable with the result estimated from point charge model of CEF. All the experimental observations indicate that  $h > J_{i j}$ in $\rm Tm_3SbO_7$, in other words, the interaction between neighboring spins is very weak. By now, we can not give the detailed pathway or strength of $J_{i j}$ from the current data.


\begin{figure}[t]
	\begin{center}
		\includegraphics[width=\columnwidth]{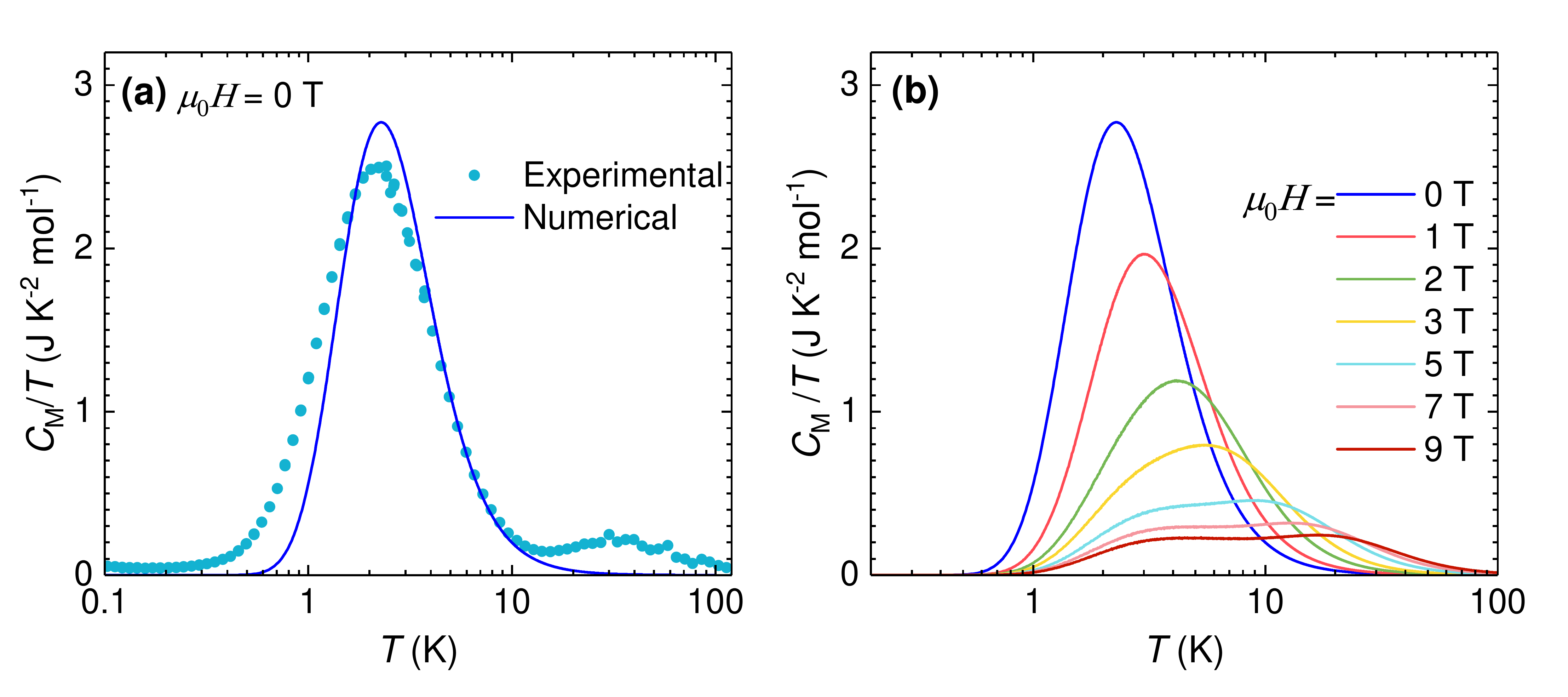}
		\caption{\textbf{(a)} Comparison between experimental results and numerical results of $C_{\rm{M}}/T$ for Tm1 under zero magnetic field. \textbf{(d)} Numerical results of $C_{\rm{M}}/T$ for Tm1 under different magnetic fields.}
		\label{fig6}
	\end{center}
\end{figure}

\subsection{\boldmath Simulation of specific heat data} \label{sec:simulation}

Based on the analysis in Sec~\ref{sec:TFIM}, a single ion model is enough to capture the principal low-temperature physics of $\rm Tm_3SbO_3$. To prove this view, we model the quasi-doublet system quantitatively and compare with the experimental data of specific heat.

Assuming the interaction between the neighboring spins is negligible, the lower-temperature bump in $C_{\rm{M}}/T$ shown in Fig. \ref{fig6}(a) should be a Schottky anomaly. For a two-level Schottky anomaly, the maximum occurs at $T_{\rm M} \approx 0.42h$. Therefore, $h$ = 7.4~K ($\sim$0.64 meV) can be derived, with the same order of magnitude of the one from CEF calculation. The single-ion Hamiltonian can be expressed explicitly as 
\begin{equation}\label{eq:single_ion_H}
	H=\frac{1}{2} h\left(\begin{array}{ll}
	0 & 1 \\
	1 & 0
	\end{array}\right)+\mu B_{0} \cos \theta\left(\begin{array}{ll}
	1 & 0 \\
	0 & -1
	\end{array}\right)
	\end{equation}

The first term in Eq.~\ref{eq:single_ion_H} is the Hamiltonian in zero field. The second term represents the effect of external magnetic field, whose magnitude is $B_0$. In a polycrystalline sample, the spins point to different directions. As a result, an external magnetic field has different effects on these spins. In our calculation, we asssum $\theta$, the angle between the easy axis of spins and the external magnetic field, is uniformly distributed in $[0,\pi]$. $\mu$ = 7.53~$\mu_{\rm{B}}$ $\approx$ 5.058~K/T is obtained from magnetic susceptibility experiments in Sec.~\ref{sec:sus}.

We simulate the specific heat by averaging over 10$^5$ random $\theta$ after obtaining each thermodynamic $C_V$ with a fixed $\theta$. A comparison between the experimental data and theoretical result in zero field is shown in Fig.~\ref{fig6}(a), which indicates that a single-ion Hamiltonian is appropriate for the system, except for a slight inconsistency at low temperatures. The theoretical results (Fig.~\ref{fig6}(b)) are also consistent with the lower-temperature bump of $C_{\rm{M}}/T$ under different external magnetic fields: the bump becomes broader and lower with field increasing. One reason is that external field expands the energy gap between the two levels, which will drive the bump to higher temperatures and thus suppress the maximum of $C_{\rm{M}}/T$. Another reason is that, in a polycrystalline sample, the distribution of $\theta$ leads to different energy splitting under a same external field. As a result, averaging all bumps with different maximum position give rise to a broader and lower bump as the external field is increased.


\section{CONCLUSIONS} \label{sec:conc}
A new 3D Sandglass magnet Tm$_3$SbO$_7$ with non-Kramers ions has been discovered. We have carried out magnetic susceptibility, specific heat, and $\mu$SR experiments on polycrystalline samples. No long-range magnetic order and no signature of spin freezing were observed down to 0.1~K. The low-energy properties of Tm$_3$SbO$_7$ are dominated by the two lowest energy levels of Tm$^{3+}_1$ with a finite energy gap, and the CEF splitting can be considered as an intrinsic transverse field. Due to the small exchange interactions between effective $S=1/2$ spins signified by the small Curie-Weiss temperature at low temperature, the TFIM with a quantum paramagnetic state can be applied. Persistent spin dynamics are observed below 1~K and at least 3~kOe. The perturbation fr0m Tm$^{3+}_2$ may play an important role to account for this dynamics behavior.

On the other hand, although geometry frustration is not found based on the lattice structure of Tm$_3$SbO$_7$, the possibility that the absence of magnetic order and the existence of spin dynamics are due to the competition of exchange interactions between the nearest-neighbor and the next-nearest-neighbor Tm$^{3+}$ ions, can not be completely excluded. 

Tm$_3$SbO$_7$ provides a new platform for studying quantum magnetism and dynamic properties. To further identify the ground state of Tm$_3$SbO$_7$, high-quality single crystals,  careful measurements of CEF splitting using inelastic neutron scattering are needed. Using pressure or element doping to regulate the energy splitting $h$ can also be revealing. 
Besides, from the point-charge-model's calculation of CEF, we also find the size of unit cell along $c$ axis have a significant influence on the several lowest energy levels of Tm2. Even 0.1~\AA ~can lead to a nearly degenerate doublet which offers a promising way to regulate the energy level and investigate how exchange interaction come into play in this system.

\begin{acknowledgments}
We thank Gang Chen, Yang Qi, Yuan Wan, and Jianda Wu for fruitful discussion. We are grateful to the ISIS cryogenics Group for their valuable help during the $\mu$SR experiments (10.5286/ISIS.E.RB1820271). This research was funded by the National Natural Science Foundations of China, No. 12034004 and 12174065, and the Shanghai Municipal Science and Technology (Major Project Grant No. 2019SHZDZX01 and No. 20ZR1405300).
\end{acknowledgments}


\begin{thebibliography}{42}%
\makeatletter
\providecommand \@ifxundefined [1]{%
 \@ifx{#1\undefined}
}%
\providecommand \@ifnum [1]{%
 \ifnum #1\expandafter \@firstoftwo
 \else \expandafter \@secondoftwo
 \fi
}%
\providecommand \@ifx [1]{%
 \ifx #1\expandafter \@firstoftwo
 \else \expandafter \@secondoftwo
 \fi
}%
\providecommand \natexlab [1]{#1}%
\providecommand \enquote  [1]{``#1''}%
\providecommand \bibnamefont  [1]{#1}%
\providecommand \bibfnamefont [1]{#1}%
\providecommand \citenamefont [1]{#1}%
\providecommand \href@noop [0]{\@secondoftwo}%
\providecommand \href [0]{\begingroup \@sanitize@url \@href}%
\providecommand \@href[1]{\@@startlink{#1}\@@href}%
\providecommand \@@href[1]{\endgroup#1\@@endlink}%
\providecommand \@sanitize@url [0]{\catcode `\\12\catcode `\$12\catcode
  `\&12\catcode `\#12\catcode `\^12\catcode `\_12\catcode `\%12\relax}%
\providecommand \@@startlink[1]{}%
\providecommand \@@endlink[0]{}%
\providecommand \url  [0]{\begingroup\@sanitize@url \@url }%
\providecommand \@url [1]{\endgroup\@href {#1}{\urlprefix }}%
\providecommand \urlprefix  [0]{URL }%
\providecommand \Eprint [0]{\href }%
\providecommand \doibase [0]{http://dx.doi.org/}%
\providecommand \selectlanguage [0]{\@gobble}%
\providecommand \bibinfo  [0]{\@secondoftwo}%
\providecommand \bibfield  [0]{\@secondoftwo}%
\providecommand \translation [1]{[#1]}%
\providecommand \BibitemOpen [0]{}%
\providecommand \bibitemStop [0]{}%
\providecommand \bibitemNoStop [0]{.\EOS\space}%
\providecommand \EOS [0]{\spacefactor3000\relax}%
\providecommand \BibitemShut  [1]{\csname bibitem#1\endcsname}%
\let\auto@bib@innerbib\@empty
\bibitem [{\citenamefont {Keimer}\ and\ \citenamefont
  {Moore}(2017)}]{Keimer2017}%
  \BibitemOpen
  \bibfield  {author} {\bibinfo {author} {\bibfnamefont {B.}~\bibnamefont
  {Keimer}}\ and\ \bibinfo {author} {\bibfnamefont {J.~E.}\ \bibnamefont
  {Moore}},\ }\href {\doibase 10.1038/nphys4302} {\bibfield  {journal}
  {\bibinfo  {journal} {Nat. Phys.}\ }\textbf {\bibinfo {volume} {13}},\
  \bibinfo {pages} {1045} (\bibinfo {year} {2017})}\BibitemShut {NoStop}%
\bibitem [{\citenamefont {Tokura}\ \emph {et~al.}(2017)\citenamefont {Tokura},
  \citenamefont {Kawasaki},\ and\ \citenamefont {Nagaosa}}]{Tokura2017}%
  \BibitemOpen
  \bibfield  {author} {\bibinfo {author} {\bibfnamefont {Y.}~\bibnamefont
  {Tokura}}, \bibinfo {author} {\bibfnamefont {M.}~\bibnamefont {Kawasaki}}, \
  and\ \bibinfo {author} {\bibfnamefont {N.}~\bibnamefont {Nagaosa}},\ }\href
  {\doibase 10.1038/nphys4274} {\bibfield  {journal} {\bibinfo  {journal} {Nat.
  Phys.}\ }\textbf {\bibinfo {volume} {13}},\ \bibinfo {pages} {1056} (\bibinfo
  {year} {2017})}\BibitemShut {NoStop}%
\bibitem [{\citenamefont {Anderson}(1973)}]{Anderson1973}%
  \BibitemOpen
  \bibfield  {author} {\bibinfo {author} {\bibfnamefont {P.~W.}\ \bibnamefont
  {Anderson}},\ }\href {\doibase https://doi.org/10.1016/0025-5408(73)90167-0}
  {\bibfield  {journal} {\bibinfo  {journal} {Mater. Res. Bull.}\ }\textbf
  {\bibinfo {volume} {8}},\ \bibinfo {pages} {153} (\bibinfo {year}
  {1973})}\BibitemShut {NoStop}%
\bibitem [{\citenamefont {Kitaev}(2006)}]{Kitaev2006}%
  \BibitemOpen
  \bibfield  {author} {\bibinfo {author} {\bibfnamefont {A.}~\bibnamefont
  {Kitaev}},\ }\href {\doibase https://doi.org/10.1016/j.aop.2005.10.005}
  {\bibfield  {journal} {\bibinfo  {journal} {Ann. Phys.}\ }\textbf {\bibinfo
  {volume} {321}},\ \bibinfo {pages} {2} (\bibinfo {year} {2006})},\ \bibinfo
  {note} {january Special Issue}\BibitemShut {NoStop}%
\bibitem [{\citenamefont {Broholm}\ \emph {et~al.}(2020)\citenamefont
  {Broholm}, \citenamefont {Cava}, \citenamefont {Kivelson}, \citenamefont
  {Nocera}, \citenamefont {Norman},\ and\ \citenamefont
  {Senthil}}]{Broholm2020}%
  \BibitemOpen
  \bibfield  {author} {\bibinfo {author} {\bibfnamefont {C.}~\bibnamefont
  {Broholm}}, \bibinfo {author} {\bibfnamefont {R.~J.}\ \bibnamefont {Cava}},
  \bibinfo {author} {\bibfnamefont {S.~A.}\ \bibnamefont {Kivelson}}, \bibinfo
  {author} {\bibfnamefont {D.~G.}\ \bibnamefont {Nocera}}, \bibinfo {author}
  {\bibfnamefont {M.~R.}\ \bibnamefont {Norman}}, \ and\ \bibinfo {author}
  {\bibfnamefont {T.}~\bibnamefont {Senthil}},\ }\href {\doibase
  10.1126/science.aay0668} {\bibfield  {journal} {\bibinfo  {journal}
  {Science}\ }\textbf {\bibinfo {volume} {367}},\ \bibinfo {pages} {eaay0668}
  (\bibinfo {year} {2020})}\BibitemShut {NoStop}%
\bibitem [{\citenamefont {Wen}\ \emph {et~al.}(2019)\citenamefont {Wen},
  \citenamefont {Yu}, \citenamefont {Li}, \citenamefont {Yu},\ and\
  \citenamefont {Li}}]{Wen2019}%
  \BibitemOpen
  \bibfield  {author} {\bibinfo {author} {\bibfnamefont {J.}~\bibnamefont
  {Wen}}, \bibinfo {author} {\bibfnamefont {S.-L.}\ \bibnamefont {Yu}},
  \bibinfo {author} {\bibfnamefont {S.}~\bibnamefont {Li}}, \bibinfo {author}
  {\bibfnamefont {W.}~\bibnamefont {Yu}}, \ and\ \bibinfo {author}
  {\bibfnamefont {J.-X.}\ \bibnamefont {Li}},\ }\href {\doibase
  10.1038/s41535-019-0151-6} {\bibfield  {journal} {\bibinfo  {journal} {npj
  Quantum Mater.}\ }\textbf {\bibinfo {volume} {4}},\ \bibinfo {pages} {12}
  (\bibinfo {year} {2019})}\BibitemShut {NoStop}%
\bibitem [{\citenamefont {Kimchi}\ \emph
  {et~al.}(2018{\natexlab{a}})\citenamefont {Kimchi}, \citenamefont
  {Sheckelton}, \citenamefont {McQueen},\ and\ \citenamefont
  {Lee}}]{Kimchi2018NC}%
  \BibitemOpen
  \bibfield  {author} {\bibinfo {author} {\bibfnamefont {I.}~\bibnamefont
  {Kimchi}}, \bibinfo {author} {\bibfnamefont {J.~P.}\ \bibnamefont
  {Sheckelton}}, \bibinfo {author} {\bibfnamefont {T.~M.}\ \bibnamefont
  {McQueen}}, \ and\ \bibinfo {author} {\bibfnamefont {P.~A.}\ \bibnamefont
  {Lee}},\ }\href {\doibase 10.1038/s41467-018-06800-2} {\bibfield  {journal}
  {\bibinfo  {journal} {Nat. Commun.}\ }\textbf {\bibinfo {volume} {9}},\
  \bibinfo {pages} {4367} (\bibinfo {year} {2018}{\natexlab{a}})}\BibitemShut
  {NoStop}%
\bibitem [{\citenamefont {Kimchi}\ \emph
  {et~al.}(2018{\natexlab{b}})\citenamefont {Kimchi}, \citenamefont {Nahum},\
  and\ \citenamefont {Senthil}}]{Kimchi2018prx}%
  \BibitemOpen
  \bibfield  {author} {\bibinfo {author} {\bibfnamefont {I.}~\bibnamefont
  {Kimchi}}, \bibinfo {author} {\bibfnamefont {A.}~\bibnamefont {Nahum}}, \
  and\ \bibinfo {author} {\bibfnamefont {T.}~\bibnamefont {Senthil}},\ }\href
  {\doibase 10.1103/PhysRevX.8.031028} {\bibfield  {journal} {\bibinfo
  {journal} {Phys. Rev. X}\ }\textbf {\bibinfo {volume} {8}},\ \bibinfo {pages}
  {031028} (\bibinfo {year} {2018}{\natexlab{b}})}\BibitemShut {NoStop}%
\bibitem [{\citenamefont {Keren}\ \emph {et~al.}(2004)\citenamefont {Keren},
  \citenamefont {Gardner}, \citenamefont {Ehlers}, \citenamefont {Fukaya},
  \citenamefont {Segal},\ and\ \citenamefont {Uemura}}]{Keren2004}%
  \BibitemOpen
  \bibfield  {author} {\bibinfo {author} {\bibfnamefont {A.}~\bibnamefont
  {Keren}}, \bibinfo {author} {\bibfnamefont {J.~S.}\ \bibnamefont {Gardner}},
  \bibinfo {author} {\bibfnamefont {G.}~\bibnamefont {Ehlers}}, \bibinfo
  {author} {\bibfnamefont {A.}~\bibnamefont {Fukaya}}, \bibinfo {author}
  {\bibfnamefont {E.}~\bibnamefont {Segal}}, \ and\ \bibinfo {author}
  {\bibfnamefont {Y.~J.}\ \bibnamefont {Uemura}},\ }\href {\doibase
  10.1103/PhysRevLett.92.107204} {\bibfield  {journal} {\bibinfo  {journal}
  {Phys. Rev. Lett.}\ }\textbf {\bibinfo {volume} {92}},\ \bibinfo {pages}
  {107204} (\bibinfo {year} {2004})}\BibitemShut {NoStop}%
\bibitem [{\citenamefont {Ueland}\ \emph {et~al.}(2006)\citenamefont {Ueland},
  \citenamefont {Lau}, \citenamefont {Cava}, \citenamefont {O'Brien},\ and\
  \citenamefont {Schiffer}}]{Ueland2006}%
  \BibitemOpen
  \bibfield  {author} {\bibinfo {author} {\bibfnamefont {B.~G.}\ \bibnamefont
  {Ueland}}, \bibinfo {author} {\bibfnamefont {G.~C.}\ \bibnamefont {Lau}},
  \bibinfo {author} {\bibfnamefont {R.~J.}\ \bibnamefont {Cava}}, \bibinfo
  {author} {\bibfnamefont {J.~R.}\ \bibnamefont {O'Brien}}, \ and\ \bibinfo
  {author} {\bibfnamefont {P.}~\bibnamefont {Schiffer}},\ }\href {\doibase
  10.1103/PhysRevLett.96.027216} {\bibfield  {journal} {\bibinfo  {journal}
  {Phys. Rev. Lett.}\ }\textbf {\bibinfo {volume} {96}},\ \bibinfo {pages}
  {027216} (\bibinfo {year} {2006})}\BibitemShut {NoStop}%
\bibitem [{\citenamefont {Ding}\ \emph {et~al.}(2018)\citenamefont {Ding},
  \citenamefont {Yang}, \citenamefont {Zhang}, \citenamefont {Tan},
  \citenamefont {Zhu}, \citenamefont {Chen},\ and\ \citenamefont
  {Shu}}]{Ding2018}%
  \BibitemOpen
  \bibfield  {author} {\bibinfo {author} {\bibfnamefont {Z.-F.}\ \bibnamefont
  {Ding}}, \bibinfo {author} {\bibfnamefont {Y.-X.}\ \bibnamefont {Yang}},
  \bibinfo {author} {\bibfnamefont {J.}~\bibnamefont {Zhang}}, \bibinfo
  {author} {\bibfnamefont {C.}~\bibnamefont {Tan}}, \bibinfo {author}
  {\bibfnamefont {Z.-H.}\ \bibnamefont {Zhu}}, \bibinfo {author} {\bibfnamefont
  {G.}~\bibnamefont {Chen}}, \ and\ \bibinfo {author} {\bibfnamefont
  {L.}~\bibnamefont {Shu}},\ }\href {\doibase 10.1103/PhysRevB.98.174404}
  {\bibfield  {journal} {\bibinfo  {journal} {Phys. Rev. B}\ }\textbf {\bibinfo
  {volume} {98}},\ \bibinfo {pages} {174404} (\bibinfo {year}
  {2018})}\BibitemShut {NoStop}%
\bibitem [{\citenamefont {Ni}\ \emph {et~al.}(2019)\citenamefont {Ni},
  \citenamefont {Pan}, \citenamefont {Song}, \citenamefont {Huang},
  \citenamefont {Zeng}, \citenamefont {Yu}, \citenamefont {Cheng},
  \citenamefont {Wang}, \citenamefont {Dai}, \citenamefont {Kato},\ and\
  \citenamefont {Li}}]{Ni2019}%
  \BibitemOpen
  \bibfield  {author} {\bibinfo {author} {\bibfnamefont {J.~M.}\ \bibnamefont
  {Ni}}, \bibinfo {author} {\bibfnamefont {B.~L.}\ \bibnamefont {Pan}},
  \bibinfo {author} {\bibfnamefont {B.~Q.}\ \bibnamefont {Song}}, \bibinfo
  {author} {\bibfnamefont {Y.~Y.}\ \bibnamefont {Huang}}, \bibinfo {author}
  {\bibfnamefont {J.~Y.}\ \bibnamefont {Zeng}}, \bibinfo {author}
  {\bibfnamefont {Y.~J.}\ \bibnamefont {Yu}}, \bibinfo {author} {\bibfnamefont
  {E.~J.}\ \bibnamefont {Cheng}}, \bibinfo {author} {\bibfnamefont {L.~S.}\
  \bibnamefont {Wang}}, \bibinfo {author} {\bibfnamefont {D.~Z.}\ \bibnamefont
  {Dai}}, \bibinfo {author} {\bibfnamefont {R.}~\bibnamefont {Kato}}, \ and\
  \bibinfo {author} {\bibfnamefont {S.~Y.}\ \bibnamefont {Li}},\ }\href
  {\doibase 10.1103/PhysRevLett.123.247204} {\bibfield  {journal} {\bibinfo
  {journal} {Phys. Rev. Lett.}\ }\textbf {\bibinfo {volume} {123}},\ \bibinfo
  {pages} {247204} (\bibinfo {year} {2019})}\BibitemShut {NoStop}%
\bibitem [{\citenamefont {Yamashita}\ \emph
  {et~al.}(2008{\natexlab{a}})\citenamefont {Yamashita}, \citenamefont
  {Nakazawa}, \citenamefont {Oguni}, \citenamefont {Oshima}, \citenamefont
  {Nojiri}, \citenamefont {Shimizu}, \citenamefont {Miyagawa},\ and\
  \citenamefont {Kanoda}}]{Satoshi2008c}%
  \BibitemOpen
  \bibfield  {author} {\bibinfo {author} {\bibfnamefont {S.}~\bibnamefont
  {Yamashita}}, \bibinfo {author} {\bibfnamefont {Y.}~\bibnamefont {Nakazawa}},
  \bibinfo {author} {\bibfnamefont {M.}~\bibnamefont {Oguni}}, \bibinfo
  {author} {\bibfnamefont {Y.}~\bibnamefont {Oshima}}, \bibinfo {author}
  {\bibfnamefont {H.}~\bibnamefont {Nojiri}}, \bibinfo {author} {\bibfnamefont
  {Y.}~\bibnamefont {Shimizu}}, \bibinfo {author} {\bibfnamefont
  {K.}~\bibnamefont {Miyagawa}}, \ and\ \bibinfo {author} {\bibfnamefont
  {K.}~\bibnamefont {Kanoda}},\ }\href {\doibase 10.1038/nphys942} {\bibfield
  {journal} {\bibinfo  {journal} {Nat. Phys.}\ }\textbf {\bibinfo {volume}
  {4}},\ \bibinfo {pages} {459} (\bibinfo {year}
  {2008}{\natexlab{a}})}\BibitemShut {NoStop}%
\bibitem [{\citenamefont {Yamashita}\ \emph {et~al.}(2011)\citenamefont
  {Yamashita}, \citenamefont {Yamamoto}, \citenamefont {Nakazawa},
  \citenamefont {Tamura},\ and\ \citenamefont {Kato}}]{Satoshi2011c}%
  \BibitemOpen
  \bibfield  {author} {\bibinfo {author} {\bibfnamefont {S.}~\bibnamefont
  {Yamashita}}, \bibinfo {author} {\bibfnamefont {T.}~\bibnamefont {Yamamoto}},
  \bibinfo {author} {\bibfnamefont {Y.}~\bibnamefont {Nakazawa}}, \bibinfo
  {author} {\bibfnamefont {M.}~\bibnamefont {Tamura}}, \ and\ \bibinfo {author}
  {\bibfnamefont {R.}~\bibnamefont {Kato}},\ }\href {\doibase
  10.1038/ncomms1274} {\bibfield  {journal} {\bibinfo  {journal} {Nat.
  Commun.}\ }\textbf {\bibinfo {volume} {2}},\ \bibinfo {pages} {275} (\bibinfo
  {year} {2011})}\BibitemShut {NoStop}%
\bibitem [{\citenamefont {Yamashita}\ \emph {et~al.}(2010)\citenamefont
  {Yamashita}, \citenamefont {Nakata}, \citenamefont {Senshu}, \citenamefont
  {Nagata}, \citenamefont {Yamamoto~Hiroshi}, \citenamefont {Kato},
  \citenamefont {Shibauchi},\ and\ \citenamefont {Matsuda}}]{Minoru2010kappa}%
  \BibitemOpen
  \bibfield  {author} {\bibinfo {author} {\bibfnamefont {M.}~\bibnamefont
  {Yamashita}}, \bibinfo {author} {\bibfnamefont {N.}~\bibnamefont {Nakata}},
  \bibinfo {author} {\bibfnamefont {Y.}~\bibnamefont {Senshu}}, \bibinfo
  {author} {\bibfnamefont {M.}~\bibnamefont {Nagata}}, \bibinfo {author}
  {\bibfnamefont {M.}~\bibnamefont {Yamamoto~Hiroshi}}, \bibinfo {author}
  {\bibfnamefont {R.}~\bibnamefont {Kato}}, \bibinfo {author} {\bibfnamefont
  {T.}~\bibnamefont {Shibauchi}}, \ and\ \bibinfo {author} {\bibfnamefont
  {Y.}~\bibnamefont {Matsuda}},\ }\href {\doibase 10.1126/science.1188200}
  {\bibfield  {journal} {\bibinfo  {journal} {Science}\ }\textbf {\bibinfo
  {volume} {328}},\ \bibinfo {pages} {1246} (\bibinfo {year}
  {2010})}\BibitemShut {NoStop}%
\bibitem [{\citenamefont {Yamashita}\ \emph
  {et~al.}(2008{\natexlab{b}})\citenamefont {Yamashita}, \citenamefont
  {Nakata}, \citenamefont {Kasahara}, \citenamefont {Sasaki}, \citenamefont
  {Yoneyama}, \citenamefont {Kobayashi}, \citenamefont {Fujimoto},
  \citenamefont {Shibauchi},\ and\ \citenamefont {Matsuda}}]{Minoru2008kappa}%
  \BibitemOpen
  \bibfield  {author} {\bibinfo {author} {\bibfnamefont {M.}~\bibnamefont
  {Yamashita}}, \bibinfo {author} {\bibfnamefont {N.}~\bibnamefont {Nakata}},
  \bibinfo {author} {\bibfnamefont {Y.}~\bibnamefont {Kasahara}}, \bibinfo
  {author} {\bibfnamefont {T.}~\bibnamefont {Sasaki}}, \bibinfo {author}
  {\bibfnamefont {N.}~\bibnamefont {Yoneyama}}, \bibinfo {author}
  {\bibfnamefont {N.}~\bibnamefont {Kobayashi}}, \bibinfo {author}
  {\bibfnamefont {S.}~\bibnamefont {Fujimoto}}, \bibinfo {author}
  {\bibfnamefont {T.}~\bibnamefont {Shibauchi}}, \ and\ \bibinfo {author}
  {\bibfnamefont {Y.}~\bibnamefont {Matsuda}},\ }\href {\doibase
  10.1038/nphys1134} {\bibfield  {journal} {\bibinfo  {journal} {Nat. Phys.}\
  }\textbf {\bibinfo {volume} {5}},\ \bibinfo {pages} {44} (\bibinfo {year}
  {2008}{\natexlab{b}})}\BibitemShut {NoStop}%
\bibitem [{\citenamefont {Coldea}\ \emph {et~al.}(2001)\citenamefont {Coldea},
  \citenamefont {Tennant}, \citenamefont {Tsvelik},\ and\ \citenamefont
  {Tylczynski}}]{Coldea2001}%
  \BibitemOpen
  \bibfield  {author} {\bibinfo {author} {\bibfnamefont {R.}~\bibnamefont
  {Coldea}}, \bibinfo {author} {\bibfnamefont {D.~A.}\ \bibnamefont {Tennant}},
  \bibinfo {author} {\bibfnamefont {A.~M.}\ \bibnamefont {Tsvelik}}, \ and\
  \bibinfo {author} {\bibfnamefont {Z.}~\bibnamefont {Tylczynski}},\ }\href
  {\doibase 10.1103/PhysRevLett.86.1335} {\bibfield  {journal} {\bibinfo
  {journal} {Phys. Rev. Lett.}\ }\textbf {\bibinfo {volume} {86}},\ \bibinfo
  {pages} {1335} (\bibinfo {year} {2001})}\BibitemShut {NoStop}%
\bibitem [{\citenamefont {Banerjee}\ \emph {et~al.}(2017)\citenamefont
  {Banerjee}, \citenamefont {Yan}, \citenamefont {Knolle}, \citenamefont
  {Bridges}, \citenamefont {Stone}, \citenamefont {Lumsden}, \citenamefont
  {Mandrus}, \citenamefont {Tennant}, \citenamefont {Moessner},\ and\
  \citenamefont {Nagler}}]{Banerjee2017}%
  \BibitemOpen
  \bibfield  {author} {\bibinfo {author} {\bibfnamefont {A.}~\bibnamefont
  {Banerjee}}, \bibinfo {author} {\bibfnamefont {J.}~\bibnamefont {Yan}},
  \bibinfo {author} {\bibfnamefont {J.}~\bibnamefont {Knolle}}, \bibinfo
  {author} {\bibfnamefont {C.~A.}\ \bibnamefont {Bridges}}, \bibinfo {author}
  {\bibfnamefont {M.~B.}\ \bibnamefont {Stone}}, \bibinfo {author}
  {\bibfnamefont {M.~D.}\ \bibnamefont {Lumsden}}, \bibinfo {author}
  {\bibfnamefont {D.~G.}\ \bibnamefont {Mandrus}}, \bibinfo {author}
  {\bibfnamefont {D.~A.}\ \bibnamefont {Tennant}}, \bibinfo {author}
  {\bibfnamefont {R.}~\bibnamefont {Moessner}}, \ and\ \bibinfo {author}
  {\bibfnamefont {S.~E.}\ \bibnamefont {Nagler}},\ }\href {\doibase
  10.1126/science.aah6015} {\bibfield  {journal} {\bibinfo  {journal}
  {Science}\ }\textbf {\bibinfo {volume} {356}},\ \bibinfo {pages} {1055}
  (\bibinfo {year} {2017})}\BibitemShut {NoStop}%
\bibitem [{\citenamefont {Dunsiger}\ \emph {et~al.}(2011)\citenamefont
  {Dunsiger}, \citenamefont {Aczel}, \citenamefont {Arguello}, \citenamefont
  {Dabkowska}, \citenamefont {Dabkowski}, \citenamefont {Du}, \citenamefont
  {Goko}, \citenamefont {Javanparast}, \citenamefont {Lin}, \citenamefont
  {Ning}, \citenamefont {Noad}, \citenamefont {Singh}, \citenamefont
  {Williams}, \citenamefont {Uemura}, \citenamefont {Gingras},\ and\
  \citenamefont {Luke}}]{Dunsiger2011}%
  \BibitemOpen
  \bibfield  {author} {\bibinfo {author} {\bibfnamefont {S.~R.}\ \bibnamefont
  {Dunsiger}}, \bibinfo {author} {\bibfnamefont {A.~A.}\ \bibnamefont {Aczel}},
  \bibinfo {author} {\bibfnamefont {C.}~\bibnamefont {Arguello}}, \bibinfo
  {author} {\bibfnamefont {H.}~\bibnamefont {Dabkowska}}, \bibinfo {author}
  {\bibfnamefont {A.}~\bibnamefont {Dabkowski}}, \bibinfo {author}
  {\bibfnamefont {M.~H.}\ \bibnamefont {Du}}, \bibinfo {author} {\bibfnamefont
  {T.}~\bibnamefont {Goko}}, \bibinfo {author} {\bibfnamefont {B.}~\bibnamefont
  {Javanparast}}, \bibinfo {author} {\bibfnamefont {T.}~\bibnamefont {Lin}},
  \bibinfo {author} {\bibfnamefont {F.~L.}\ \bibnamefont {Ning}}, \bibinfo
  {author} {\bibfnamefont {H.~M.}\ \bibnamefont {Noad}}, \bibinfo {author}
  {\bibfnamefont {D.~J.}\ \bibnamefont {Singh}}, \bibinfo {author}
  {\bibfnamefont {T.~J.}\ \bibnamefont {Williams}}, \bibinfo {author}
  {\bibfnamefont {Y.~J.}\ \bibnamefont {Uemura}}, \bibinfo {author}
  {\bibfnamefont {M.~J.}\ \bibnamefont {Gingras}}, \ and\ \bibinfo {author}
  {\bibfnamefont {G.~M.}\ \bibnamefont {Luke}},\ }\href {\doibase
  10.1103/PhysRevLett.107.207207} {\bibfield  {journal} {\bibinfo  {journal}
  {Phys. Rev. Lett.}\ }\textbf {\bibinfo {volume} {107}},\ \bibinfo {pages}
  {207207} (\bibinfo {year} {2011})}\BibitemShut {NoStop}%
\bibitem [{\citenamefont {Chang}\ \emph {et~al.}(2013)\citenamefont {Chang},
  \citenamefont {Lees}, \citenamefont {Balakrishnan}, \citenamefont {Kao},\
  and\ \citenamefont {Hillier}}]{Chang2013}%
  \BibitemOpen
  \bibfield  {author} {\bibinfo {author} {\bibfnamefont {L.~J.}\ \bibnamefont
  {Chang}}, \bibinfo {author} {\bibfnamefont {M.~R.}\ \bibnamefont {Lees}},
  \bibinfo {author} {\bibfnamefont {G.}~\bibnamefont {Balakrishnan}}, \bibinfo
  {author} {\bibfnamefont {Y.~J.}\ \bibnamefont {Kao}}, \ and\ \bibinfo
  {author} {\bibfnamefont {A.~D.}\ \bibnamefont {Hillier}},\ }\href {\doibase
  10.1038/srep01881} {\bibfield  {journal} {\bibinfo  {journal} {Sci. Rep.}\
  }\textbf {\bibinfo {volume} {3}},\ \bibinfo {pages} {1881} (\bibinfo {year}
  {2013})}\BibitemShut {NoStop}%
\bibitem [{\citenamefont {Balents}(2010)}]{Balents2010}%
  \BibitemOpen
  \bibfield  {author} {\bibinfo {author} {\bibfnamefont {L.}~\bibnamefont
  {Balents}},\ }\href {\doibase 10.1038/nature08917} {\bibfield  {journal}
  {\bibinfo  {journal} {Nature}\ }\textbf {\bibinfo {volume} {464}},\ \bibinfo
  {pages} {199} (\bibinfo {year} {2010})}\BibitemShut {NoStop}%
\bibitem [{\citenamefont {Zhou}\ \emph {et~al.}(2017)\citenamefont {Zhou},
  \citenamefont {Kanoda},\ and\ \citenamefont {Ng}}]{ZhouYi2017}%
  \BibitemOpen
  \bibfield  {author} {\bibinfo {author} {\bibfnamefont {Y.}~\bibnamefont
  {Zhou}}, \bibinfo {author} {\bibfnamefont {K.}~\bibnamefont {Kanoda}}, \ and\
  \bibinfo {author} {\bibfnamefont {T.-K.}\ \bibnamefont {Ng}},\ }\href
  {\doibase 10.1103/RevModPhys.89.025003} {\bibfield  {journal} {\bibinfo
  {journal} {Rev. Mod. Phys.}\ }\textbf {\bibinfo {volume} {89}},\ \bibinfo
  {pages} {025003} (\bibinfo {year} {2017})}\BibitemShut {NoStop}%
\bibitem [{\citenamefont {Kitagawa}\ \emph {et~al.}(2018)\citenamefont
  {Kitagawa}, \citenamefont {Takayama}, \citenamefont {Matsumoto},
  \citenamefont {Kato}, \citenamefont {Takano}, \citenamefont {Kishimoto},
  \citenamefont {Bette}, \citenamefont {Dinnebier}, \citenamefont {Jackeli},\
  and\ \citenamefont {Takagi}}]{Kitagawa2018}%
  \BibitemOpen
  \bibfield  {author} {\bibinfo {author} {\bibfnamefont {K.}~\bibnamefont
  {Kitagawa}}, \bibinfo {author} {\bibfnamefont {T.}~\bibnamefont {Takayama}},
  \bibinfo {author} {\bibfnamefont {Y.}~\bibnamefont {Matsumoto}}, \bibinfo
  {author} {\bibfnamefont {A.}~\bibnamefont {Kato}}, \bibinfo {author}
  {\bibfnamefont {R.}~\bibnamefont {Takano}}, \bibinfo {author} {\bibfnamefont
  {Y.}~\bibnamefont {Kishimoto}}, \bibinfo {author} {\bibfnamefont
  {S.}~\bibnamefont {Bette}}, \bibinfo {author} {\bibfnamefont
  {R.}~\bibnamefont {Dinnebier}}, \bibinfo {author} {\bibfnamefont
  {G.}~\bibnamefont {Jackeli}}, \ and\ \bibinfo {author} {\bibfnamefont
  {H.}~\bibnamefont {Takagi}},\ }\href {\doibase 10.1038/nature25482}
  {\bibfield  {journal} {\bibinfo  {journal} {Nature}\ }\textbf {\bibinfo
  {volume} {554}},\ \bibinfo {pages} {341} (\bibinfo {year}
  {2018})}\BibitemShut {NoStop}%
\bibitem [{\citenamefont {Pei}\ \emph {et~al.}(2020)\citenamefont {Pei},
  \citenamefont {Huang}, \citenamefont {Li}, \citenamefont {Chen},
  \citenamefont {Xi}, \citenamefont {Wang}, \citenamefont {Shi}, \citenamefont
  {Yu}, \citenamefont {Liu}, \citenamefont {Wang}, \citenamefont {Ye},
  \citenamefont {Huang},\ and\ \citenamefont {Mei}}]{Pei2020}%
  \BibitemOpen
  \bibfield  {author} {\bibinfo {author} {\bibfnamefont {S.}~\bibnamefont
  {Pei}}, \bibinfo {author} {\bibfnamefont {L.-L.}\ \bibnamefont {Huang}},
  \bibinfo {author} {\bibfnamefont {G.}~\bibnamefont {Li}}, \bibinfo {author}
  {\bibfnamefont {X.}~\bibnamefont {Chen}}, \bibinfo {author} {\bibfnamefont
  {B.}~\bibnamefont {Xi}}, \bibinfo {author} {\bibfnamefont {X.}~\bibnamefont
  {Wang}}, \bibinfo {author} {\bibfnamefont {Y.}~\bibnamefont {Shi}}, \bibinfo
  {author} {\bibfnamefont {D.}~\bibnamefont {Yu}}, \bibinfo {author}
  {\bibfnamefont {C.}~\bibnamefont {Liu}}, \bibinfo {author} {\bibfnamefont
  {L.}~\bibnamefont {Wang}}, \bibinfo {author} {\bibfnamefont {F.}~\bibnamefont
  {Ye}}, \bibinfo {author} {\bibfnamefont {M.}~\bibnamefont {Huang}}, \ and\
  \bibinfo {author} {\bibfnamefont {J.-W.}\ \bibnamefont {Mei}},\ }\href
  {\doibase 10.1103/PhysRevB.101.201101} {\bibfield  {journal} {\bibinfo
  {journal} {Phys. Rev. B}\ }\textbf {\bibinfo {volume} {101}},\ \bibinfo
  {pages} {201101} (\bibinfo {year} {2020})}\BibitemShut {NoStop}%
\bibitem [{\citenamefont {Okamoto}\ \emph {et~al.}(2007)\citenamefont
  {Okamoto}, \citenamefont {Nohara}, \citenamefont {Aruga-Katori},\ and\
  \citenamefont {Takagi}}]{Okamoto2007}%
  \BibitemOpen
  \bibfield  {author} {\bibinfo {author} {\bibfnamefont {Y.}~\bibnamefont
  {Okamoto}}, \bibinfo {author} {\bibfnamefont {M.}~\bibnamefont {Nohara}},
  \bibinfo {author} {\bibfnamefont {H.}~\bibnamefont {Aruga-Katori}}, \ and\
  \bibinfo {author} {\bibfnamefont {H.}~\bibnamefont {Takagi}},\ }\href
  {\doibase 10.1103/PhysRevLett.99.137207} {\bibfield  {journal} {\bibinfo
  {journal} {Phys. Rev. Lett.}\ }\textbf {\bibinfo {volume} {99}},\ \bibinfo
  {pages} {137207} (\bibinfo {year} {2007})}\BibitemShut {NoStop}%
\bibitem [{\citenamefont {Chen}\ and\ \citenamefont
  {Balents}(2008)}]{ChenGang2008}%
  \BibitemOpen
  \bibfield  {author} {\bibinfo {author} {\bibfnamefont {G.}~\bibnamefont
  {Chen}}\ and\ \bibinfo {author} {\bibfnamefont {L.}~\bibnamefont {Balents}},\
  }\href {\doibase 10.1103/PhysRevB.78.094403} {\bibfield  {journal} {\bibinfo
  {journal} {Phys. Rev. B}\ }\textbf {\bibinfo {volume} {78}},\ \bibinfo
  {pages} {094403} (\bibinfo {year} {2008})}\BibitemShut {NoStop}%
\bibitem [{\citenamefont {Zhou}\ \emph {et~al.}(2008)\citenamefont {Zhou},
  \citenamefont {Lee}, \citenamefont {Ng},\ and\ \citenamefont
  {Zhang}}]{Zhou2008}%
  \BibitemOpen
  \bibfield  {author} {\bibinfo {author} {\bibfnamefont {Y.}~\bibnamefont
  {Zhou}}, \bibinfo {author} {\bibfnamefont {P.~A.}\ \bibnamefont {Lee}},
  \bibinfo {author} {\bibfnamefont {T.-K.}\ \bibnamefont {Ng}}, \ and\ \bibinfo
  {author} {\bibfnamefont {F.-C.}\ \bibnamefont {Zhang}},\ }\href {\doibase
  10.1103/PhysRevLett.101.197201} {\bibfield  {journal} {\bibinfo  {journal}
  {Phys. Rev. Lett.}\ }\textbf {\bibinfo {volume} {101}},\ \bibinfo {pages}
  {197201} (\bibinfo {year} {2008})}\BibitemShut {NoStop}%
\bibitem [{\citenamefont {Nakatsuji}\ \emph {et~al.}(2006)\citenamefont
  {Nakatsuji}, \citenamefont {Machida}, \citenamefont {Maeno}, \citenamefont
  {Tayama}, \citenamefont {Sakakibara}, \citenamefont {Duijn}, \citenamefont
  {Balicas}, \citenamefont {Millican}, \citenamefont {Macaluso},\ and\
  \citenamefont {Chan}}]{Nakatsuji2006}%
  \BibitemOpen
  \bibfield  {author} {\bibinfo {author} {\bibfnamefont {S.}~\bibnamefont
  {Nakatsuji}}, \bibinfo {author} {\bibfnamefont {Y.}~\bibnamefont {Machida}},
  \bibinfo {author} {\bibfnamefont {Y.}~\bibnamefont {Maeno}}, \bibinfo
  {author} {\bibfnamefont {T.}~\bibnamefont {Tayama}}, \bibinfo {author}
  {\bibfnamefont {T.}~\bibnamefont {Sakakibara}}, \bibinfo {author}
  {\bibfnamefont {J.}~\bibnamefont {Duijn}}, \bibinfo {author} {\bibfnamefont
  {L.}~\bibnamefont {Balicas}}, \bibinfo {author} {\bibfnamefont {J.~N.}\
  \bibnamefont {Millican}}, \bibinfo {author} {\bibfnamefont {R.~T.}\
  \bibnamefont {Macaluso}}, \ and\ \bibinfo {author} {\bibfnamefont {J.~Y.}\
  \bibnamefont {Chan}},\ }\href {\doibase 10.1103/PhysRevLett.96.087204}
  {\bibfield  {journal} {\bibinfo  {journal} {Phys. Rev. Lett.}\ }\textbf
  {\bibinfo {volume} {96}},\ \bibinfo {pages} {087204} (\bibinfo {year}
  {2006})}\BibitemShut {NoStop}%
\bibitem [{\citenamefont {Chen}(2016)}]{ChenGang2016}%
  \BibitemOpen
  \bibfield  {author} {\bibinfo {author} {\bibfnamefont {G.}~\bibnamefont
  {Chen}},\ }\href {\doibase 10.1103/PhysRevB.94.205107} {\bibfield  {journal}
  {\bibinfo  {journal} {Phys. Rev. B}\ }\textbf {\bibinfo {volume} {94}},\
  \bibinfo {pages} {205107} (\bibinfo {year} {2016})}\BibitemShut {NoStop}%
\bibitem [{\citenamefont {Yao}\ and\ \citenamefont {Chen}(2018)}]{Yao2018}%
  \BibitemOpen
  \bibfield  {author} {\bibinfo {author} {\bibfnamefont {X.-P.}\ \bibnamefont
  {Yao}}\ and\ \bibinfo {author} {\bibfnamefont {G.}~\bibnamefont {Chen}},\
  }\href {\doibase 10.1103/PhysRevX.8.041039} {\bibfield  {journal} {\bibinfo
  {journal} {Phys. Rev. X}\ }\textbf {\bibinfo {volume} {8}},\ \bibinfo {pages}
  {041039} (\bibinfo {year} {2018})}\BibitemShut {NoStop}%
\bibitem [{\citenamefont {Ni}\ \emph {et~al.}(2021)\citenamefont {Ni},
  \citenamefont {Huang}, \citenamefont {Cheng}, \citenamefont {Yu},
  \citenamefont {Pan}, \citenamefont {Li}, \citenamefont {Xu}, \citenamefont
  {Tian},\ and\ \citenamefont {Li}}]{Ni2021}%
  \BibitemOpen
  \bibfield  {author} {\bibinfo {author} {\bibfnamefont {J.~M.}\ \bibnamefont
  {Ni}}, \bibinfo {author} {\bibfnamefont {Y.~Y.}\ \bibnamefont {Huang}},
  \bibinfo {author} {\bibfnamefont {E.~J.}\ \bibnamefont {Cheng}}, \bibinfo
  {author} {\bibfnamefont {Y.~J.}\ \bibnamefont {Yu}}, \bibinfo {author}
  {\bibfnamefont {B.~L.}\ \bibnamefont {Pan}}, \bibinfo {author} {\bibfnamefont
  {Q.}~\bibnamefont {Li}}, \bibinfo {author} {\bibfnamefont {L.~M.}\
  \bibnamefont {Xu}}, \bibinfo {author} {\bibfnamefont {Z.~M.}\ \bibnamefont
  {Tian}}, \ and\ \bibinfo {author} {\bibfnamefont {S.~Y.}\ \bibnamefont
  {Li}},\ }\href {\doibase 10.1038/s41467-020-20562-w} {\bibfield  {journal}
  {\bibinfo  {journal} {Nat. Commun.}\ }\textbf {\bibinfo {volume} {12}},\
  \bibinfo {pages} {307} (\bibinfo {year} {2021})}\BibitemShut {NoStop}%
\bibitem [{\citenamefont {Coldea}\ \emph {et~al.}(2010)\citenamefont {Coldea},
  \citenamefont {Tennant}, \citenamefont {Wheeler}, \citenamefont {Wawrzynska},
  \citenamefont {Prabhakaran}, \citenamefont {Telling}, \citenamefont
  {Habicht}, \citenamefont {Smeibidl},\ and\ \citenamefont
  {Kiefer}}]{Coldea2010}%
  \BibitemOpen
  \bibfield  {author} {\bibinfo {author} {\bibfnamefont {R.}~\bibnamefont
  {Coldea}}, \bibinfo {author} {\bibfnamefont {D.~A.}\ \bibnamefont {Tennant}},
  \bibinfo {author} {\bibfnamefont {E.~M.}\ \bibnamefont {Wheeler}}, \bibinfo
  {author} {\bibfnamefont {E.}~\bibnamefont {Wawrzynska}}, \bibinfo {author}
  {\bibfnamefont {D.}~\bibnamefont {Prabhakaran}}, \bibinfo {author}
  {\bibfnamefont {M.}~\bibnamefont {Telling}}, \bibinfo {author} {\bibfnamefont
  {K.}~\bibnamefont {Habicht}}, \bibinfo {author} {\bibfnamefont
  {P.}~\bibnamefont {Smeibidl}}, \ and\ \bibinfo {author} {\bibfnamefont
  {K.}~\bibnamefont {Kiefer}},\ }\href {\doibase 10.1126/science.1180085}
  {\bibfield  {journal} {\bibinfo  {journal} {Science}\ }\textbf {\bibinfo
  {volume} {327}},\ \bibinfo {pages} {177} (\bibinfo {year}
  {2010})}\BibitemShut {NoStop}%
\bibitem [{\citenamefont {Shen}\ \emph {et~al.}(2019)\citenamefont {Shen},
  \citenamefont {Liu}, \citenamefont {Qin}, \citenamefont {Shen}, \citenamefont
  {Li}, \citenamefont {Bewley}, \citenamefont {Schneidewind}, \citenamefont
  {Chen},\ and\ \citenamefont {Zhao}}]{ShenYao2019}%
  \BibitemOpen
  \bibfield  {author} {\bibinfo {author} {\bibfnamefont {Y.}~\bibnamefont
  {Shen}}, \bibinfo {author} {\bibfnamefont {C.}~\bibnamefont {Liu}}, \bibinfo
  {author} {\bibfnamefont {Y.}~\bibnamefont {Qin}}, \bibinfo {author}
  {\bibfnamefont {S.}~\bibnamefont {Shen}}, \bibinfo {author} {\bibfnamefont
  {Y.~D.}\ \bibnamefont {Li}}, \bibinfo {author} {\bibfnamefont
  {R.}~\bibnamefont {Bewley}}, \bibinfo {author} {\bibfnamefont
  {A.}~\bibnamefont {Schneidewind}}, \bibinfo {author} {\bibfnamefont
  {G.}~\bibnamefont {Chen}}, \ and\ \bibinfo {author} {\bibfnamefont
  {J.}~\bibnamefont {Zhao}},\ }\href {\doibase 10.1038/s41467-019-12410-3}
  {\bibfield  {journal} {\bibinfo  {journal} {Nat. Commun.}\ }\textbf {\bibinfo
  {volume} {10}},\ \bibinfo {pages} {4530} (\bibinfo {year}
  {2019})}\BibitemShut {NoStop}%
\bibitem [{\citenamefont {Cui}\ \emph {et~al.}(2019)\citenamefont {Cui},
  \citenamefont {Zou}, \citenamefont {Xi}, \citenamefont {He}, \citenamefont
  {Yang}, \citenamefont {Shu}, \citenamefont {Zhang}, \citenamefont {Hu},
  \citenamefont {Chen}, \citenamefont {Yu}, \citenamefont {Wu},\ and\
  \citenamefont {Yu}}]{Cui2019}%
  \BibitemOpen
  \bibfield  {author} {\bibinfo {author} {\bibfnamefont {Y.}~\bibnamefont
  {Cui}}, \bibinfo {author} {\bibfnamefont {H.}~\bibnamefont {Zou}}, \bibinfo
  {author} {\bibfnamefont {N.}~\bibnamefont {Xi}}, \bibinfo {author}
  {\bibfnamefont {Z.}~\bibnamefont {He}}, \bibinfo {author} {\bibfnamefont
  {Y.~X.}\ \bibnamefont {Yang}}, \bibinfo {author} {\bibfnamefont
  {L.}~\bibnamefont {Shu}}, \bibinfo {author} {\bibfnamefont {G.~H.}\
  \bibnamefont {Zhang}}, \bibinfo {author} {\bibfnamefont {Z.}~\bibnamefont
  {Hu}}, \bibinfo {author} {\bibfnamefont {T.}~\bibnamefont {Chen}}, \bibinfo
  {author} {\bibfnamefont {R.}~\bibnamefont {Yu}}, \bibinfo {author}
  {\bibfnamefont {J.}~\bibnamefont {Wu}}, \ and\ \bibinfo {author}
  {\bibfnamefont {W.}~\bibnamefont {Yu}},\ }\href {\doibase
  10.1103/PhysRevLett.123.067203} {\bibfield  {journal} {\bibinfo  {journal}
  {Phys. Rev. Lett.}\ }\textbf {\bibinfo {volume} {123}},\ \bibinfo {pages}
  {067203} (\bibinfo {year} {2019})}\BibitemShut {NoStop}%
\bibitem [{\citenamefont {Liu}\ \emph {et~al.}(2020)\citenamefont {Liu},
  \citenamefont {Huang},\ and\ \citenamefont {Chen}}]{Liu2020}%
  \BibitemOpen
  \bibfield  {author} {\bibinfo {author} {\bibfnamefont {C.}~\bibnamefont
  {Liu}}, \bibinfo {author} {\bibfnamefont {C.-J.}\ \bibnamefont {Huang}}, \
  and\ \bibinfo {author} {\bibfnamefont {G.}~\bibnamefont {Chen}},\ }\href
  {\doibase 10.1103/PhysRevResearch.2.043013} {\bibfield  {journal} {\bibinfo
  {journal} {Phys. Rev. Res.}\ }\textbf {\bibinfo {volume} {2}},\ \bibinfo
  {pages} {043013} (\bibinfo {year} {2020})}\BibitemShut {NoStop}%
\bibitem [{\citenamefont {Li}\ \emph {et~al.}(2020)\citenamefont {Li},
  \citenamefont {Liao}, \citenamefont {Chen}, \citenamefont {Zeng},
  \citenamefont {Sheng}, \citenamefont {Qi}, \citenamefont {Meng},\ and\
  \citenamefont {Li}}]{LiHan2020}%
  \BibitemOpen
  \bibfield  {author} {\bibinfo {author} {\bibfnamefont {H.}~\bibnamefont
  {Li}}, \bibinfo {author} {\bibfnamefont {Y.~D.}\ \bibnamefont {Liao}},
  \bibinfo {author} {\bibfnamefont {B.~B.}\ \bibnamefont {Chen}}, \bibinfo
  {author} {\bibfnamefont {X.~T.}\ \bibnamefont {Zeng}}, \bibinfo {author}
  {\bibfnamefont {X.~L.}\ \bibnamefont {Sheng}}, \bibinfo {author}
  {\bibfnamefont {Y.}~\bibnamefont {Qi}}, \bibinfo {author} {\bibfnamefont
  {Z.~Y.}\ \bibnamefont {Meng}}, \ and\ \bibinfo {author} {\bibfnamefont
  {W.}~\bibnamefont {Li}},\ }\href {\doibase 10.1038/s41467-020-14907-8}
  {\bibfield  {journal} {\bibinfo  {journal} {Nat. Commun.}\ }\textbf {\bibinfo
  {volume} {11}},\ \bibinfo {pages} {1111} (\bibinfo {year}
  {2020})}\BibitemShut {NoStop}%
\bibitem [{\citenamefont {Chen}(2019)}]{ChenGang2019}%
  \BibitemOpen
  \bibfield  {author} {\bibinfo {author} {\bibfnamefont {G.}~\bibnamefont
  {Chen}},\ }\href {\doibase 10.1103/PhysRevResearch.1.033141} {\bibfield
  {journal} {\bibinfo  {journal} {Phys. Rev. Res.}\ }\textbf {\bibinfo {volume}
  {1}},\ \bibinfo {pages} {033141} (\bibinfo {year} {2019})}\BibitemShut
  {NoStop}%
\bibitem [{\citenamefont {Bitko}\ \emph {et~al.}(1996)\citenamefont {Bitko},
  \citenamefont {Rosenbaum},\ and\ \citenamefont {Aeppli}}]{Bitko1996}%
  \BibitemOpen
  \bibfield  {author} {\bibinfo {author} {\bibfnamefont {D.}~\bibnamefont
  {Bitko}}, \bibinfo {author} {\bibfnamefont {T.~F.}\ \bibnamefont
  {Rosenbaum}}, \ and\ \bibinfo {author} {\bibfnamefont {G.}~\bibnamefont
  {Aeppli}},\ }\href {\doibase 10.1103/PhysRevLett.77.940} {\bibfield
  {journal} {\bibinfo  {journal} {Phys. Rev. Lett.}\ }\textbf {\bibinfo
  {volume} {77}},\ \bibinfo {pages} {940} (\bibinfo {year} {1996})}\BibitemShut
  {NoStop}%
\bibitem [{\citenamefont {Inabayashi}\ \emph {et~al.}(2018)\citenamefont
  {Inabayashi}, \citenamefont {Doi}, \citenamefont {Wakeshima},\ and\
  \citenamefont {Hinatsu}}]{Masaki2018}%
  \BibitemOpen
  \bibfield  {author} {\bibinfo {author} {\bibfnamefont {M.}~\bibnamefont
  {Inabayashi}}, \bibinfo {author} {\bibfnamefont {Y.}~\bibnamefont {Doi}},
  \bibinfo {author} {\bibfnamefont {M.}~\bibnamefont {Wakeshima}}, \ and\
  \bibinfo {author} {\bibfnamefont {Y.}~\bibnamefont {Hinatsu}},\ }\href
  {\doibase 10.2109/jcersj2.18127} {\bibfield  {journal} {\bibinfo  {journal}
  {J. Ceram. Soc. JAPAN}\ }\textbf {\bibinfo {volume} {126}},\ \bibinfo {pages}
  {920} (\bibinfo {year} {2018})}\BibitemShut {NoStop}%
\bibitem [{\citenamefont {Scheie}(2021)}]{PyCrystalField}%
  \BibitemOpen
  \bibfield  {author} {\bibinfo {author} {\bibfnamefont {A.}~\bibnamefont
  {Scheie}},\ }\href {\doibase 10.1107/s160057672001554x} {\bibfield  {journal}
  {\bibinfo  {journal} {J. Appl. Crystallogr.}\ }\textbf {\bibinfo {volume}
  {54}},\ \bibinfo {pages} {356} (\bibinfo {year} {2021})}\BibitemShut
  {NoStop}%
\bibitem [{\citenamefont {Hayano}\ \emph {et~al.}(1979)\citenamefont {Hayano},
  \citenamefont {Uemura}, \citenamefont {Imazato}, \citenamefont {Nishida},
  \citenamefont {Yamazaki},\ and\ \citenamefont {Kubo}}]{Hayano79}%
  \BibitemOpen
  \bibfield  {author} {\bibinfo {author} {\bibfnamefont {R.~S.}\ \bibnamefont
  {Hayano}}, \bibinfo {author} {\bibfnamefont {Y.~J.}\ \bibnamefont {Uemura}},
  \bibinfo {author} {\bibfnamefont {J.}~\bibnamefont {Imazato}}, \bibinfo
  {author} {\bibfnamefont {N.}~\bibnamefont {Nishida}}, \bibinfo {author}
  {\bibfnamefont {T.}~\bibnamefont {Yamazaki}}, \ and\ \bibinfo {author}
  {\bibfnamefont {R.}~\bibnamefont {Kubo}},\ }\href {\doibase
  10.1103/PhysRevB.20.850} {\bibfield  {journal} {\bibinfo  {journal} {Phys.
  Rev. B}\ }\textbf {\bibinfo {volume} {20}},\ \bibinfo {pages} {850} (\bibinfo
  {year} {1979})}\BibitemShut {NoStop}%
\bibitem [{\citenamefont {MacLaughlin}\ \emph {et~al.}(2009)\citenamefont
  {MacLaughlin}, \citenamefont {Ohta}, \citenamefont {Machida}, \citenamefont
  {Nakatsuji}, \citenamefont {Luke}, \citenamefont {Ishida}, \citenamefont
  {Heffner}, \citenamefont {Shu},\ and\ \citenamefont
  {Bernal}}]{MacLaughlin2009}%
  \BibitemOpen
  \bibfield  {author} {\bibinfo {author} {\bibfnamefont {D.~E.}\ \bibnamefont
  {MacLaughlin}}, \bibinfo {author} {\bibfnamefont {Y.}~\bibnamefont {Ohta}},
  \bibinfo {author} {\bibfnamefont {Y.}~\bibnamefont {Machida}}, \bibinfo
  {author} {\bibfnamefont {S.}~\bibnamefont {Nakatsuji}}, \bibinfo {author}
  {\bibfnamefont {G.~M.}\ \bibnamefont {Luke}}, \bibinfo {author}
  {\bibfnamefont {K.}~\bibnamefont {Ishida}}, \bibinfo {author} {\bibfnamefont
  {R.~H.}\ \bibnamefont {Heffner}}, \bibinfo {author} {\bibfnamefont
  {L.}~\bibnamefont {Shu}}, \ and\ \bibinfo {author} {\bibfnamefont {O.~O.}\
  \bibnamefont {Bernal}},\ }\href {\doibase 10.1016/j.physb.2008.11.167}
  {\bibfield  {journal} {\bibinfo  {journal} {Physica B}\ }\textbf {\bibinfo
  {volume} {404}},\ \bibinfo {pages} {667} (\bibinfo {year}
  {2009})}\BibitemShut {NoStop}%
\end{thebibliography}
%

\end{document}